\documentclass[preprint2]{aastex}

\shorttitle{Unstable neutrino disk}
\shortauthors{Janiuk et al.}

\begin{document}

\title{Instabilities in the time-dependent neutrino disc in Gamma-Ray Bursts}

\author{A. Janiuk\altaffilmark{1,2}, Y. Yuan\altaffilmark{3},
  R. Perna\altaffilmark{4} \&  T. Di Matteo\altaffilmark{5}}
\altaffiltext{1}{Copernicus Astronomical Center, Bartycka 18, 00-716 Warsaw, 
Poland}
\altaffiltext{2} {Present address: University of Nevada, Las Vegas, 4505 Maryland Pkwy, NV89154, USA}
\altaffiltext{3}{ Center for Astrophysics, University of Science and Technology of China, Hefei, Anhui 230026, P.R. China}
\altaffiltext{4}{JILA and Department of Astrophysical and Planetary Sciences, University of Colorado, 440 UCB, Boulder, CO 80309, USA}
\altaffiltext{5}{Physics Department, Carnegie Mellon University, 5000 Forbes Avenue, Pittsburgh, PA 15232, USA}

\begin{abstract}
We investigate the properties and evolution of accretion tori formed
after the
coalescence of two compact objects. At these extreme densities and
temperatures, the accreting torus is cooled mainly by neutrino
emission produced primarily by electron and positron capture on
nucleons ($\beta$ reactions). We solve for the disc structure and its
time evolution by introducing a detailed treatment of the equation of
state which includes photodisintegration of helium, the condition of
$\beta$-equilibrium, and neutrino opacities.  We self-consistently
calculate the chemical equilibrium in the gas consisting of helium,
free protons, neutrons and electron-positron pairs and compute the
chemical potentials of the species, as well as the electron fraction
throughout the disc. We find that, for sufficiently large accretion
rates ($\dot M \ga 10 M_{\odot}$/s), the inner regions of the disk
become opaque and develop a viscous and thermal instability.  The
identification of this instability 
might be relevant for GRB observations.
\end{abstract}


\section{Introduction}
Gamma-Ray Bursts are commonly thought to be produced in relativistic
ejecta that dissipate energy by internal shocks 
however alternative ideas based on the Poynting flux dominated
  jets are also being proposed (for a review see e.g. Piran 2005;
  Meszaros 2006; Zhang 2007).

The enormous power released during the
gamma-ray burst explosion indicates that a relativistic phenomenon
must be involved in creating GRBs (Narayan et al. 1992).  The merger
of two neutron stars or of a neutron star and a black hole (e.g. NS-NS
or NS-BH) has been invoked e.g. by Paczy\'nski (1986); Eichler et al. (1989); Paczy\'nski
(1991); Narayan, Paczy\'nski \& Piran (1992), as well as the collapse
of a massive star, the so called ``collapsar'' scenario (e.g.,~Woosley
1993 and Paczy\'nski 1998).  In both cases a dense, hot accretion disk
is likely to form around a newly born black hole (Witt et al. 1994).
In the collapsar scenario, the collapsing envelope of the star
accretes onto the newly formed black hole, while a transient debris
disk is formed when the NS-NS or NS-BH binary merges (see e.g. Ruffert
et al. 1997 for numerical simulations).

The durations of GRBs, which range from milliseconds to over a thousand of
seconds, are distributed in two distinct peaks defining two main GRB
classes: short ($\le 2$ sec) and long ($\ga 2$ sec) bursts (Kouvelietou et
al. 1993).  For some long bursts, signatures of an accompanying supernova
explosion have been detected in the afterglow spectra (Stanek et al. 2003; Hjorth et al. 2003),
which strongly favors the ``collapsar'' interpretation for their
origin. Furthermore, the GRB positions inferred from the afterglow
observations are consistent with the GRBs being associated with the star
forming regions in their host galaxies. For short bursts, several pieces of
evidence from the analysis of the {\em Swift} satellite and follow-up observations
(Gehrels et al. 2005; Fox et al. 2005; Villasenor et al. 2005) argue in favor
of a binary merger model (Hjorth et al. 2005; Berger et al. 2005).

In the merger scenario, the duration of the burst is comparable to
the viscous timescale of the accretion disc whereas in the collapsar scenario,
the external reservoir of stellar matter can feed the accretion torus for a
much longer timescale.  In general, accretion discs powering GRBs are expected
to have typical densities of the order of $10^{10-12}$ g cm$^{-3}$ and
temperatures of $10^{11}$ K within $10-20$ Schwarzschild radii ($R_{\rm S}
= 2GM/c^2$).  Thus, accretion proceeds with rates of a fraction to 
several solar masses per second. 
In this ``hyper-accreting'' regime, photons become trapped
and are not efficient at cooling the disc.  Neutrinos, however, are produced
by weak interactions in the dense and hot plasma, releasing the
gravitational energy of the accretion flow. These discs go under the name of
``neutrino-dominated accretion flows", or NDAFs.  

Over the last several years a number of studies have investigated the structure
of these discs (Popham, Woosley \& Fryer 1999; Narayan, Piran \& Kumar 2001;
Kohri \& Mineshige 2002; Di Matteo, Perna \& Narayan 2002; Surman \&
McLaughlin 2004; Kohri, Narayan \& Piran 2005; Chen \& Beloborodov 2006; 
Gu, Liu \& Lu 2006; Liu, Gu \& Lu 2007).
These models have employed the customary approximation of one-dimensional
hydrodynamics (Shakura \& Sunyaev 1973), where the effects of MHD viscous
stresses are described by the dimensionless parameter $\alpha$, but have been
limited to the steady-state approximation of constant $\dot{M}$.  Such an
assumption is a good approximation when considering the collapsar scenario,
where the burst duration is much longer than the viscous time, due to the
continuous replenishing of the disc by the collapsing star envelope.  However,
even in this scenario, recent observations suggest that that engines are
``long-lived'' (past the torus feeding phase), requiring a time-dependent
computation. In the merger scenario, on the other hand, a time-dependent
calculation is necessary even for modeling the prompt phase of the burst, since the
duration is set by the viscous timescale of the disc.

Recently, a fully time-dependent calculation of the structure of such
accretion discs has been presented by Janiuk et al. (2004). The model was
suitable for the torus being a results of either the gravitational 
collapse of a massive stellar core or
 the compact binary merger. However the
structure and evolution of the disk (like in many of the early
calculations) was calculated under a number of simplifying assumptions for the
composition and the equation of state of the accreting matter.
In this paper we improve upon our earlier results in several ways.
Besides the requirement of time-dependent calculations, the high density and
temperature regime in which the accreting gas lies, implies that both
multi-dimensional numerical and semi-analytic calculations for such flows need
to include the detailed microphysics. This includes photodisintegration of
nuclei, the establishment of statistical equilibrium, neutronization, and
the effects of neutrino opacities in the inner regions. 
Here, we introduce a detailed treatment of the equation of state,
and calculate self-consistently the chemical equilibrium in the gas that
consists of helium, free protons, neutrons and electron-positron pairs.  We
compute the chemical potentials of the species, as well as the electron
fraction throughout the disk using the assumption of the equilibrium between
the beta processes. Our EOS equations include, self-consistently, the
contribution of the neutrino trapping to the beta equilibrium. Another
important addition compared to our previous work (Janiuk et al. 2004) 
is the inclusion of 
photodisintegration of helium. 
The presence of this term can affect the energy balance in the 
inner, opaque (to neutrinos as well
as photons) region of the flow and, as it will be shown, 
it eventually produces a thermal and viscous instability in those regions. 
This is especially relevant since
the GRB phenomenology requires a variable energy output.

Other time-dependent disc studies of binary mergers or collapsars have
been performed in 2D using hydrodynamical simulations (e.g. MacFadyen
\& Woosley 1999; Ruffert \& Janka 1999; Lee et al. 2002; Rosswog et
al. 2004; Lee, Ramirez-Ruiz \& Page 2005) and, most recently, in 3-D
simulations (Setiawan et al. 2005).  Also, MHD simulations of the GRB
central engine have been performed, showing that the magnetic field
possibly plays an important role in the generation of a GRB jet (Proga
et al. 2003; Fujimoto et al. 2006).  The advantage of our calculations
is that, whilst including all the relevant physics to calculate the
equation of state, the structure and stability of the accretion disc,
we are able to study a much larger range of parameter space and allow
our calculations to evolve beyond what can be reached in higher
dimensional calculation and comparable to at least the short-burst
durations.

The paper is organized as follows. In Section \ref{sec:method} we describe the
basic assumptions of the model and the method used in the initial stationary
and subsequent time-dependent numerical simulations. In Section
\ref{sec:results} we discuss the structure of the hyperaccreting disc for
various values of the initial accretion rate, and study the time evolution of
its density and temperature, as well as the resulting neutrino lightcurve.  We
also discuss the physical origin of the instabilities in the disc, and we
compare our model and results with the recent 2D and 3D simulations.  We
summarize our results in Section \ref{sec:diss}.

\section{Neutrino-cooled accretion disks}
\label{sec:method}
In this Section, we describe how we improve upon our previous
time-dependent calculation (Janiuk et al. 2004) by computing
self-consistently the equation of state of the extremely dense matter
by solving the balance of the $\beta$ reaction rates.  This allows
us to determine the chemical potentials of electrons, protons and
neutrons, as well as the electron fraction, in the initial disc
configuration and throughout its evolution.

\subsection{Initial disc configuration: 1-D Hydrodynamics}
\label{sec:eos}
We start by considering a steady-state model of an accretion disc
around a Schwarzschild black hole - formed as a remnant
structure either after a compact binary merger, or in a collapsar
after the birth of a black hole (for a recent calculation in Kerr
 spacetime see Chen \& Beloborodov 2006).
Throughout our calculations we use the vertically integrated equations
and hence derive a vertically averaged disc structure. 
We write the surface density of the disk as $\Sigma = H \rho$,
 where
$\rho$ is the density and where the disk half thickness (or disk height)
is given by $H=c_{s}/\Omega_{K}$. Here the sound speed is defined by
$c_{s} = \sqrt{P/\rho}$ and  $\Omega_{\rm K} = \sqrt{GM/r^{3}}$
is the Keplerian angular velocity with $P$ the total pressure.
We note that, at very high accretion rates, the
disc becomes moderately geometrically thick ($H \sim 0.5 r$) in regions
where neutrino cooling becomes inefficient and advection dominates. 
Our 'slim disk' approximation neglects terms $\sim (H/r)^2$ and assumes
that the fluid is in Keplerian rotation.
For the disc viscous stress we use the standard $\alpha$ viscosity 
prescription of  Shakura \& Sunyaev (1973) where the stress 
tensor is proportional to the pressure:
\begin{equation}
\tau_{r\varphi}=-\alpha P.
\end{equation}
We adopt a value of $\alpha = 0.1$

We set the inner radius of the disc  at 3~$R_{\rm S}$, while the outer 
radius is at 50~$R_{\rm S}$. The initial mass of such a disc is
about $0.35 M_{\odot}$ for an accretion rate $\dot M = 1
~M_{\odot}$/s. Throughout the calculations we adopt a black hole mass of $M=3 M_{\odot}$.

\subsection{The equation of state}
We  assume that the torus consists of helium, electron-positron pairs, free
neutrons and  protons. 
The total pressure is contributed by all particle species in the disc, and the 
fraction of  each species is determined by self-consistently solving the
balance of the beta reaction rates.
In the equation of state we take into account the pressure 
due to the free nuclei and pairs, helium, radiation and the trapped 
neutrinos: 
\begin{equation}
P = P_{\rm nucl}+P_{\rm He}+P_{\rm rad}+P_{\nu}\;.
\end{equation} 
The component $P_{\rm nucl}$ includes free neutrons, protons, 
and the electron-positron pair gas in beta equilibrium:
\begin{equation}
P_{\rm nucl}=P_{\rm e-}+P_{\rm e+}+P_{\rm n}+P_{\rm p}
\end{equation}
with
\begin{equation}
P_{\rm i} = {2 \sqrt{2}\over 3\pi^{2}}
{(m_{i}c^{2})^{4} \over (\hbar c)^{3}}\beta_{i}^{5/2}
\left[F_{3/2}(\eta_{\rm i},\beta_{\rm i})+{1\over 2} \beta_{\rm i}F_{5/2}(\eta_{\rm i},\beta_{\rm i})\right]\;, 
\label{eq:pi}
\end{equation}

\noindent where $F_{\rm k}$ are the Fermi-Dirac integrals of the order $k$, and
$\eta_{\rm e}$, $\eta_{\rm p}$ and $\eta_{\rm n}$ are the reduced chemical
potentials of electrons, protons and neutrons in units of $kT$,
respectively (where $\eta_i = \mu_i/kT$, also known as the degeneracy
parameter, where $\mu_i$ the standard chemical
potential) calculated from the chemical equilibrium condition
(\S~\ref{sec:chem_eq}). The reduced chemical potential of positrons is
$\eta_{\rm e+}=-\eta_{\rm e}-2/\beta_{\rm e}$ and the relativity parameters 
of the species $i$ are defined as $\beta_{\rm i}=kT/m_{\rm i}c^{2}$.

Under the physical
 conditions in the torus, helium is generally non-relativistic
and non-degenerate; therefore, its pressure is given by:
\begin{equation}
P_{\rm He}= n_{\rm He}kT, 
\end{equation}
where 
$n_{\rm He}$ is the number density of helium. This is defined as:
\begin{equation}
n_{\rm He}={1\over 4}n_{b}(1-X_{\rm nuc})\;,
\end{equation}
and the fraction of free nucleons is given by
\begin{equation}
X_{\rm nuc}=295.5\rho_{10}^{-3/4}T_{11}^{9/8}\exp(-0.8209/T_{11}),
\label{eq:xnuc}
\end{equation}
with $T_{11}$ the temperature in unit of 10$^{11}$ K (e.g. Qian \&
Woosley 1996; Popham et al. 1999). 

The radiation pressure is given by:
\begin{equation}
P_{\rm rad}={1\over 3}{\pi^{2}\over 15}{(kT)^{4}\over (\hbar c)^{3}} . 
\end{equation}

When neutrinos become trapped in the disc, the neutrino pressure is non-zero. 
Following the treatment of photon transport under the
two-stream approximation (Popham \& Narayan 1995; Di Matteo et al. 2002), 
we have
\begin{eqnarray}
P_{\nu}&=& {7\over8}{\pi^{2}\over 15}{(kT)^{4}\over3(\hbar c)^{3}} 
\sum_{i=e,\mu,\tau} {{1\over 2}(\tau_{a, \nu_{i}}+ \tau_{s}) + {1\over \sqrt 3} \over 
{1\over 2} (\tau_{a, \nu_{i}}+ \tau_{s}) + {1\over \sqrt 3} + {1 \over 3 \tau_{a, \nu_{i}}}} \nonumber \\
&\equiv & {7\over8}{\pi^{2}\over 15}{(kT)^{4}\over3(\hbar c)^{3}}b,
\label{eq:pnu}
\end{eqnarray}
where $\tau_{\rm s}$ is the scattering optical depth due to the neutrino 
scattering on
free neutrons and protons and $\tau_{\rm a, \nu_{e}}$ and $\tau_{\rm a, \nu_{\mu}}$ are 
the absorptive optical depths for electron and muon neutrinos, respectively
(see\S 2.3).
The contribution from tau neutrinos is the same as that from muon neutrinos.
These optical depths and neutrino absorption processes (which are the reverse 
of the emission processes) are discussed in more detail in the Appendix.

In the disc we have to consider both the neutrino transparent and opaque regions,
as well as the transition between the two. In the transparent case,
the neutrinos are not thermalized and the chemical potential of neutrinos is negligible.  
On the other hand, when neutrinos are totally trapped, the chemical equilibrium condition
yields: $\mu_{\rm e} + \mu_{\rm p} = \mu_{\rm n} + \mu_{\nu}$. The chemical 
potential of neutrinos is a parameter depending on how much neutrinos and anti-neutrinos 
are trapped, and assuming that the number densities of the trapped neutrinos and anti-neutrinos 
are the same, $\mu_{\nu}$ can be set to zero.
In order to determine the distribution function of the partially trapped neutrinos,
in principle one should solve the Boltzmann equation. To simplify this problem, we use here
a "gray body" model, and we introduce a blocking factor  $b=\sum_{i=e,\mu,\tau} b_i$ 
to describe the extent to which neutrinos are trapped (see e.g. Sawyer 2003). In terms of this factor, we write 
the distribution function of neutrinos as
\begin{equation}
\widetilde{f}_{\nu_i}(p)=\frac{b_i}{\exp(pc/kT)+1} =b_if_{\nu_i},\,\,\,(0 \leq b_i \leq 1).
\end{equation}
This simplified assumption is consistent with the two-stream approximation
which we adopt here (Eq. \ref{eq:pnu}).

\subsection{Composition and chemical equilibrium}
\label{sec:chem_eq}
The equilibrium state of the gas in the accreting torus is
completely determined by the chemical potentials of neutrons,
protons and electrons ($\eta_n, \eta_p, \eta_e$), and the trapping
factor of neutrinos ($b$) which is related to the optical depths
of neutrinos (cf. Eq.~9). 
For a given baryon number density, $n_{\rm b}$,
temperature $T$, and a value for accretion rate $\dot{M}$ and viscous constant
$\alpha$, the chemical potentials, or equivalently  
the ratio of free protons $x=n_{p}/n_{b}$, are determined from the 
condition of equilibrium between the transition reactions from 
neutrons to protons and from protons to neutrons. These reactions 
are:
\begin{eqnarray}
\label{eq:ur1}
p + e^{-} \to n + \nu_{\rm e} \\
\label{eq:ur4}
p + \bar\nu_{\rm e} \to n + e^{+} \\
\label{eq:ur5}
p + e^{-} + \bar\nu_{e} \to n \\
\label{eq:ur2}
n + e^{+} \to p + \bar\nu_{\rm e} \\
\label{eq:ur3}
n \to p + e^{-} + \bar\nu_{\rm e} \\
\label{eq:ur6}
n + \nu_{\rm e} \to p + e^{-} 
\end{eqnarray}
Therefore we have to calculate the ratio of protons
that will satisfy the balance:
\begin{eqnarray}
&&n_{\rm p} (\Gamma_{p + e^{-} \to n + \nu_{\rm e}} + 
	\Gamma_{p + \bar\nu_{\rm e} \to n + e^{+}} +
	\Gamma_{p + e^{-} + \bar\nu_{\rm e} \to n}) 
\nonumber \\
&&=n_{\rm n} (\Gamma_{n + e^{+} \to p + \bar\nu_{\rm e}}+ 
	\Gamma_{n \to p + e^{-} + \nu_{\rm e}}+ 
	\Gamma_{n + \nu_{\rm e} \to p + e^{-}})\;.
\end{eqnarray}
The reaction rates are the sum of forward and backward rates and are given 
in the Appendix (see also Kohri, Narayan and Piran 2005).

These are supplemented by two additional conditions: 
the conservation of the baryon number, 
$n_n+n_p=n_b \times X_{\rm nuc}$, and charge neutrality (Yuan 2005):
\begin{equation}
n_{\rm e} = n_{e^{-}} - n_{e^{+}} = n_{\rm p}+n_{\rm e}^{0}\;,
\end{equation}
which says that the net number of electrons is equal to the number of free protons plus the number of protons in helium:
\begin{equation}
n_{\rm e}^{0} = 2 n_{\rm He} = (1-X_{\rm nuc}){n_{\rm b} \over 2}.
\end{equation}

The number density of fermions under arbitrary degeneracy  
is determined by the following equations:
\begin{equation}
n_{\rm i} = {\sqrt{2} \over \pi^{2}}\left({m_{i}c^{2} \over \hbar c}\right)^{3}\beta_{\rm i}^{3/2} 
\left[F_{1/2}(\eta_{\rm i},\beta_{\rm i})+\beta_{\rm i} F_{3/2}(\eta_{\rm i},\beta_{\rm i})\right].
\end{equation}

Finally, the electron fraction is defined as:
\begin{equation}
Y_{\rm e} = {n_{\rm e^{-}}-n_{\rm e^{+}} \over n_{\rm b}}
\end{equation}
(Note that this is different from $Y_{\rm e} = 1/(1+n_{\rm n}/n_{\rm p})$,
which is only valid for free n-p-e gas.) 

\subsection{Neutrino cooling}
\label{sec:neu_rate}
The processes that are responsible for the neutrino emission in the disc are
electron-positron pair annihilation ($e^{-}+e^{+}\to \nu_{\rm i}+\bar\nu_{\rm i}$),
bremsstrahlung ($n+n \to n+n+\nu_{\rm i}+\bar\nu_{\rm i}$), plasmon decay
($\tilde \gamma \to \nu_{\rm e}+\bar\nu_{\rm e}$) and URCA process (reactions \ref{eq:ur1},
\ref{eq:ur2} and \ref{eq:ur3}). 
The first two processes produce neutrinos of all flavors, while the 
other produce only electron neutrinos and anti-neutrinos.

The cooling rate due to pair annihilation is expressed as:
\begin{equation}
q_{\rm e^{+}e^{-}} = q_{\nu_{\rm e}}+q_{\nu_{\mu}}+q_{\nu_{\tau}}
\end{equation}
where the cooling rates for all three neutrino flavors are calculated by
means of Fermi-Dirac integrals and are given in the Appendix. 

The cooling rate due to nucleon-nucleon bremsstrahlung (in erg/cm$^{3}$/s)
is given by:
\begin{equation}
q_{\rm brems} = 3.35\times 10^{27} \rho_{10}^{2} T_{11}^{5.5}\;, 
\end{equation}
where $\rho_{10}$ is the baryon density in units of 10$^{10}$ g/cm$^{3}$ and
$T_{11}$ is temperature in units of 10$^{11}$ K.

The cooling rate due to the plasmon decay (in erg/cm$^{3}$/s) is:
\begin{equation}
q_{\rm plasmon} = 1.5\times 10^{32} T_{11}^{9} \gamma_{p}^{6} e^{-\gamma_{p}}
(1+\gamma_{p})\left(2+{\gamma_{p}^{2} \over 1+\gamma_{p}}\right)\;, 
\end{equation}
where $\gamma_{p}=5.565\times 10^{-2}\sqrt{(\pi^{2}+3(\mu_{e}/kT)^{2})/3}$.

The cooling rate due to the URCA reactions is given by the three emissivities:
\begin{equation}
q_{\rm urca} = q_{p+e^{-}\to n+\nu_{\rm e}} + q_{\rm n+e^{+}\to p+\bar\nu_{\rm e}} + q_{n\to p+e^{-}+\bar\nu_{\rm e}}\;.
\end{equation}
The emissivities are given in the Appendix.

Note that the blocking factor of the trapping neutrinos 
is used only for the emissivities of the URCA reactions. For simplicity,
we neglect the blocking effects of neutrinos when calculating the 
emissivities for the electron-positron pair annihilation. Two reasons make
this approximation reasonable: the 
emissivities for the electron-positron pair annihilation is much smaller than 
those of the URCA reactions, and the electron-positron pair 
annihilation does not change the electron fraction which 
sensitively affects the EOS. 

Each of the above neutrino emission process has a reverse process, which 
leads to neutrino absorption. These are given by Equations \ref{eq:ur4},
\ref{eq:ur5} and \ref{eq:ur6}.
 Therefore we introduce the absorptive 
optical depths for neutrinos given by:
\begin{equation}
\tau_{\rm a,\nu_{i}} = {H \over 4 {7 \over 8}\sigma T^{4}} q_{\rm a, \nu_{i}}
\end{equation}
where absorption of the electron neutrinos is determined by:
\begin{equation}
q_{\rm a, \nu_{e}} = q_{\nu_{e}}^{\rm pair} + q_{\rm urca} + q_{\rm plasm} + {1\over 3} q_{\rm brems}
\end{equation}
and for the muon neutrinos:
\begin{equation}
q_{\rm a, \nu_{\mu}} = q_{\nu_{\mu}}^{\rm pair} + {1\over 3} q_{\rm brems}\;.
\end{equation}

In addition, the free escape of neutrinos from the disc is limited by 
scattering.
The scattering optical depth is given by:
\begin{eqnarray}
\tau_{\rm s} &=& \tau_{\rm s,p}+\tau_{\rm s,n}  \\
&=& 24.28\times 10^{-5}
\nonumber \left[\left({kT\over m_{\rm e}c^{2}}\right)^{2} H \left(C_{\rm s,p} n_{\rm p} + C_{\rm s,n} n_{\rm n}\right)\right]
\end{eqnarray}
where $C_{\rm s,p}=(4(C_{V}-1)^{2}+5\alpha^{2})/24$,  $C_{\rm s,n}=(1+5\alpha^{2})/24$,
$C_{\rm V}=1/2+2 \sin^{2}\theta_{\rm C}$, with $\alpha=1.25$ and $\sin^{2}\theta_{\rm C}=0.23$.

The neutrino cooling rate is then given by
\begin{equation}
Q^{-}_{\nu} = { {7 \over 8} \sigma T^{4} \over 
{3 \over 4}} \sum_{i=e,\mu} { 1 \over {\tau_{\rm a, \nu_{i}} + \tau_{\rm s} \over 2} 
+ {1 \over \sqrt 3} + 
{1 \over 3\tau_{\rm a, \nu_{i}}}}\;.
\label{eq:qnuthick}
\end{equation}

\subsection{Energy and momentum Conservation}
The hydrodynamic equations we solve to calculate the disc structure are
the standard mass, energy and momentum conservation.

Making use of the standard disk equations, the vertically integrated viscous
heating rate (per unit area) over a half thickness $ H$ is given by:
\begin{equation} 
F_{\rm tot} = {3 G M \dot M \over 8 \pi r^3} f(r)
\label{eq:ftot} 
\end{equation}
 where the Newtonian boundary condition is assumed:
$f(r) = 1-\sqrt{r_{min}/r}$.
Note that in the time-dependent calculations, instead of Eq. \ref{eq:ftot}, we will solve
the viscous diffusion equation (Eq.~\ref{eq:timevo}).

Using mass and momentum conservation $\dot M = 4 \pi \rho R H v_r \approx
6 \pi \nu \rho H$ where $v_r \approx (3\nu) /(2r)$ and 
$\nu = (2 P \alpha) /
(3 \rho \Omega)$ is the kinematic viscosity.
The  viscous heating rate can be written in terms of $\alpha$,
\begin{equation}
Q^{+}_{\rm visc}={3 \over 2}\alpha \Omega H P.
\end{equation}

Cooling in the disc is due to advection, radiation and neutrino
emission. The advective cooling in a stationary disc is determined 
approximately as:
\begin{equation}
Q^{-}_{\rm adv}=  \Sigma v_{r} T {d S \over d r} = 
q_{\rm adv}{\alpha P H T \over \Omega \rho r^{2}} S\;,
\label{eq:fadv}
\end{equation}
 where $q_{\rm adv} \propto d\ln S/d \ln r \propto(d \ln T/d \ln r - (\Gamma_{3}-1)d \ln \rho/d \ln r)
  \approx const$ and we adopt the value of 1.0. 
The entropy density $S$ is the sum of four components: 
\begin{equation}
S = S_{\rm nucl}+S_{\rm He}+S_{\rm rad}+S_{\nu}\;.
\end{equation} 
The entropy density of the gas of free protons, neutrons and electron-positron pairs
is given by:
\begin{equation}
S_{\rm nucl}= S_{\rm e^{-}}+S_{\rm e^{+}}+S_{\rm p}+S_{\rm n}
\end{equation}
where
\begin{equation}
\frac{S_{\rm i}}{k} = {1\over kT}(\epsilon_{i}+P_{\rm i})-n_{\rm i}\eta_{\rm i}
\end{equation}
and
\begin{equation}
\epsilon_{i} = {2 \sqrt{2}\over 3\pi^{2}}
{(m_{\rm i}c^{2})^{4} \over (\hbar c)^{3}}\beta_{\rm i}^{5/2}
\left[F_{3/2}(\eta_{\rm i},\beta_{\rm i})+\beta_{\rm i}F_{5/2}(\eta_{\rm i},\beta_{\rm i})\right]
\end{equation}
is the energy density of electrons, positrons, protons or neutrons, 
$P_{\rm i}$ is their pressure, given by Equation~\ref{eq:pi},
 $n_{\rm i}$ are the number densities and $\eta_{\rm i}$ are the 
chemical potentials.

The entropy density of helium is given by:
\begin{equation}
S_{\rm He} = n_{\rm He} \left[{5 \over 2} + {3\over 2} \log (m_{\rm He} {kT\over m_{\rm e}c^{2}}{1\over 2\pi}-\log n_{\rm He})\right]\;,
\end{equation}
for $n_{\rm He}>0$.

The entropy density of radiation is
\begin{equation}
S_{\rm rad} = 4{P_{\rm rad} \over kT}\;,
\end{equation}
while for neutrinos we have
\begin{equation}
S_{\nu} = 4{P_{\nu} \over kT}.
\end{equation}

In the initial disc configuration we assume 
that $q_{\rm adv}$ is approximately constant and of order of unity,
but in the subsequent time-dependent evolution the advection term 
is calculated with the appropriate radial derivatives.

For the case of photon and electron-positron pairs in the plasma
the radiative cooling is equal to:
\begin{equation}
Q^{-}_{\rm rad}={3 P_{\rm rad} c \over 4\tau}={11 \sigma T^{4} \over 4
\kappa \Sigma}
\label{eq:qrad}
\end{equation}
where we adopt the Rosseland-mean opacity
$\kappa=0.4+0.64\times 10^{23}\rho T^{-3}$ ~[cm$^{2}$g$^{-1}$]. 

An important term in the cooling and heating balance 
in the disc is due to photodisintegration of $\alpha$ particles, with rate:
\begin{equation}
Q_{\rm photo} = q_{\rm photo} H
\end{equation}   
where
\begin{equation}
q_{\rm photo}= 6.28 \times 10^{28} \rho_{10} v_{r} {dX_{\rm nuc} \over dr}
\label{eq:photodis}
\end{equation}
and $X_{\rm nuc}$ is given by Equation~\ref{eq:xnuc}.
Finally, in order to calculate  the initial stationary  configuration, 
we solve the energy balance:
$F_{\rm tot} = Q^{+}_{\rm visc} = Q^{-}_{\rm adv}+Q^{-}_{\rm
  rad}+Q^{-}_{\nu} + Q_{\rm photo}$.

\subsection{Time evolution}
\label{sec:evol}
After solving for the initial disc configuration,
we allow the density and temperature to vary with time. We solve 
the time-dependent
equations of mass and angular momentum conservation in the disc:
\begin{equation}
{\partial \Sigma \over \partial t}={1 \over r}{\partial \over \partial
r}\left[3 r^{1/2} {\partial \over \partial r}(r^{1/2} \nu \Sigma)\right]
\label{eq:timevo}
\end{equation}
and the energy equation:
\begin{eqnarray}
{\partial T \over \partial t} + v_{\rm r}{\partial T \over \partial r}
= {T \over \Sigma}{4-3\chi \over 12-9\chi}\left({\partial \Sigma \over
\partial t}+  v_{\rm r}{\partial \Sigma \over \partial r}\right) \\
\nonumber +{T\over P H}{1\over 12-9\chi} (Q^{+}-Q^{-}).
\label{eq:timevol}
\end{eqnarray}
where $\chi=(P-P_{\rm rad})/P$.
The cooling term $Q^{-}$ consists
of radiative and neutrino cooling, given by Equations (\ref{eq:qrad}) and
(\ref{eq:qnuthick}).  Advection is included in 
the energy equation via the radial derivatives. 
The cooling term due to photodisintegration of helium now
must be proportional to the {\it full} time derivative of $X_{\rm nuc}$ (cf. Eq. \ref{eq:photodis}) :
\begin{equation}
Q_{\rm photo} \propto v_{\rm r}{\partial X_{nuc} \over \partial r} +
{\partial X_{\rm nuc} \over \partial t}.
\label{eq:xnucevol}
\end{equation} 

\subsection{Numerical method}
\label{sec:numeric}
The initial configuration of the disc is calculated by means of the
Newton-Raphson method, iterated with the hydrostatic equilibrium condition.
We
interpolate over the matrix of pre-calculated results for the equation
of state (pressure and entropy) and neutrino cooling rate ( the
  number of points is 1024x1024). 
 The
Fermi-Dirac integrals are calculated using the mixture of
Gauss-Legendre and Gauss-Laguerre quadratures (Aparicio 1998).

Having determined the initial radial profiles of density and
temperature, as well as the other quantities at time $t=0$, we start the
time evolution of the disc.  We solve the set of
Equations~(\ref{eq:timevo}), (\ref{eq:timevol}) and (\ref{eq:xnucevol}) using the
convenient change of variables $y=2r^{1/2}$ and $\Xi = y \Sigma$, at
fixed radial grid, equally spaced in $y$ (see Janiuk et al. 2002
and references therein).  The number of radial zones is set to
200,  which we found to be an adequate resolution.
After determining the solutions for the first 100 time steps by
the fourth-order Runge-Kutta method, we use the Adams-Moulton
predictor-corrector method with 
 an adaptive time step.
 The code used an explicit communication model that is implemented with the
standardized MPI communication interface and can be run on 
multiprocessor machines.

We choose the no-torque  inner boundary condition, $\Sigma_{\rm in} =
T_{\rm in} = 0$ (see Abramowicz \& Kato 1989). 
The outer boundary of the disc 
 is done by adding an extra ``dead-zone'' to the computational
  domain, which accounts for the disk expansion and conservation of
  angular momentum.

\section{Results}
\label{sec:results}
We first analyze the pressure, entropy and neutrino cooling rate
distributions for a given temperature and baryon density in the gas.
Then, we show the disc structure for a converged static disc model 
and finally we show examples of time evolution of the neutrino luminosity,
density, temperature and electron fraction for 
given sets of parameters.

\subsection{EOS solutions for a given temperature and density}
\begin{figure}
\epsscale{.80}
\plotone{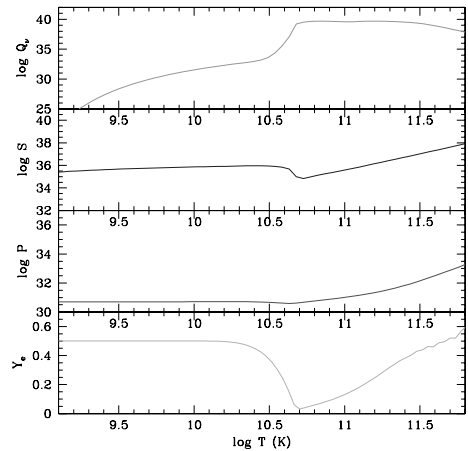}
\caption{The dependence of the electron fraction (bottom), 
pressure (middle bottom), entropy (middle upper)
 neutrino cooling rate (upper panel)
on temperature, for the constant density $\rho = 10^{12}$ g/cm$^{3}$.
The accretion rate is
$\dot M=1~ M_{\odot}$/s.
The pressure and neutrino cooling are in cgs units and the entropy is in units of 
$k_{\rm B}$ cm$^{-3}$.
\label{fig:eos1}}
\end{figure}

\begin{figure}
\epsscale{.80}
\plotone{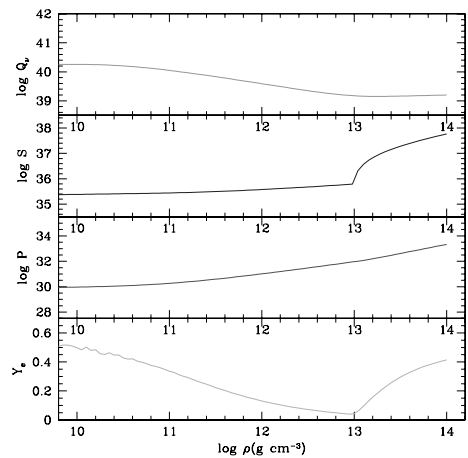}
\caption{The dependence of the electron fraction (bottom),
pressure (middle bottom), entropy (middle upper)
 and neutrino cooling rate (upper panel)
on density, for the constant temperature $T = 10^{11}$ K.
The accretion rate is
$\dot M=1~ M_{\odot}$/s.
The pressure and neutrino cooling are in cgs units and the entropy is in units of 
$k_{\rm B} cm^{-3}$.
\label{fig:eos2}}
\end{figure}

In Figure \ref{fig:eos1} and \ref{fig:eos2} we plot the results of 
the numerically calculated 
equation of state for the hot and dense matter. The plots 
 show the dependence of the electron fraction, 
pressure, entropy and neutrino cooling rate 
on temperature and density, respectively.

In the upper panels, we show the neutrino cooling
rate.
At low temperatures, below $T=m_{\rm e}c^{2}\sim 5\times 10^{9} K$,
there are almost no positrons and free nucleons. Therefore the 
neutrino emission processes switch off, and the cooling of
the gas is either due to advection, or,
when the matter becomes transparent to photons, radiative cooling overtakes. 
For larger temperatures, the neutrino emission rate
 increases up to the temperature of about $\sim 5 \times
10^{11}$ K. For very high temperatures, the optical depths for neutrinos
increase very rapidly ($\tau \propto T^{5}$, 
 see Eq. (7) in Di Matteo et al. 2002).
Therefore the neutrino cooling rate decreases at high temperatures 
(Eq. \ref{eq:qnuthick}).
On the other hand, for a given temperature (e.g. $T \sim
10^{11}$ K), the neutrino cooling rate does not sensitively depend on density.
It varies by one
 order of magnitude in the range of $10^{8} \le \rho \le 10^{14}$
g/cm$^{3}$, where the optical depth is $\tau \sim 100$.
The middle panels of Figures \ref{fig:eos1} and \ref{fig:eos2}, show
the entropy and pressure as a function of temperature and density.  At
low temperatures, the entropy of gas is not important.  The highly
degenerate electrons do not give contribution to the entropy, while
they are a dominant term in the pressure, which is therefore
independent of temperature up to $T \sim 5\times 10^{10}$ K. When the
temperature increases, helium becomes disintegrated into free
nucleons at
energy comparable to the binding energy of helium, 
and after that the radiation (including photons and
electron-positron pairs) contributes mainly to the total entropy and
pressure. Therefore, both these quantities rise with temperature.  At
high densities, the entropy is dominated by neutrons.  
Finally, the electron fraction is shown in the bottom panel of Figures 1 and
2.
At low temperatures, the electron fraction is equal to 0.5,
it then decreases sharply as the helium nuclei become disintegrated.
As the temperature further increases, positrons appear as the electrons become
non-degenerate. 
The positron capture again increases the electron fraction (see Fig.\ref{fig:eos1}).

The electron fraction changes significantly 
as a function of density for $T > 10^{10}K$ (in Fig.\ref{fig:eos2}, 
$T = 10^{11} K$). 
At low densities,
the torus consists of free neutrons and protons and
 $Y_{\rm e}$ is close to 0.5 (see also Eq.~7). 
As density increases, $Y_{\rm e}$ decreases
to satisfy the beta-equilibrium among the free n-p-e gas.
Above some density (when the temperature is high enough, e.g. for
Fig. \ref{fig:eos2}, $\rho_{\rm He}\approx10^{13}$g cm$^{-3}$) helium
starts forming. Therefore $Y_{\rm e}$ has a kink and stars rising
steeply, asymptotically approaching 0.5 as the torus consists
of plenty of ionized helium and some
electrons to keep charge neutrality.

\subsection{The  steady-state disc structure}
\label{sec:stabil}
\begin{figure}
\epsscale{.80}
\plotone{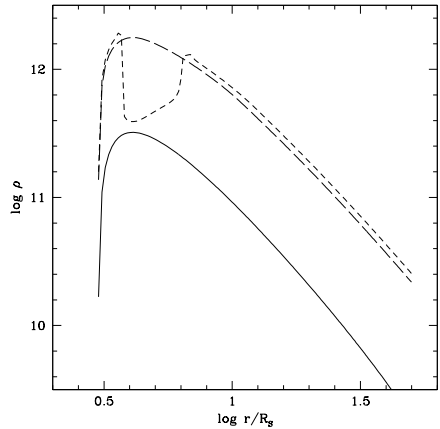}
\caption{The baryon density as a function of the disc radius,
calculated in the stationary solution.
The accretion rate is
$\dot M=1~ M_{\odot}$/s (solid line),
 $\dot M=10~ M_{\odot}$/s (long dashed line) 
  and $\dot M=12~M_{\odot}$/s (short dashed line) .
\label{fig:rho}}
\end{figure}

\begin{figure}
\epsscale{.80}
\plotone{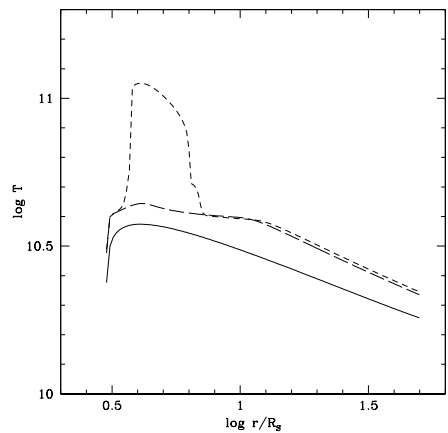}
\caption{The temperature as a function of the disc radius,
calculated in the stationary solution.
The accretion rate is
$\dot M=1~ M_{\odot}$/s (solid line),
 $\dot M=10~ M_{\odot}$/s (long dashed line)
  and $\dot M=12~M_{\odot}$/s (short dashed line) .
\label{fig:temp}}
\end{figure}

In Figures \ref{fig:rho} and \ref{fig:temp} we show the profiles of density and
 temperature in the stationary accretion disk model for three accretion rates: 
$1~ M_{\odot}$/s,  $10~ M_{\odot}$/s and $12~ M_{\odot}$/s. 
In general, the temperature and density profiles
both increase inward. 
However, for $\dot{M}=12 M_{\odot}$/s, 
a distinct branch of solutions is reached, which appears different
than the so-called ``NDAF"
branch (see Kohri \& Mineshige 2002). 
The density and temperature profiles for this high accretion rate
differ also from what was found in previous work
(Di Matteo et al. 2002; Janiuk et al. 2004). Due to a more
detailed equation of state, in which we allow for a partial
degeneracy of nucleons and electrons as well as neutrino trapping, our
solutions reach densities as high as  $10^{12}$ g/cm$^{3}$ in the innermost radii of the disc. The temperature in this
inner disc part is in the range $4 \times 10^{10}-1.25 \times 10^{11}$ K, depending on the 
accretion rate. For the hottest disk model, 
a local peak in the density forms around 
$7-8 R_{\rm S}$, while below that radius the density decreases.
Between $\sim$ 3.5 and 7 $R_{\rm S}$,
the plasma becomes much hotter and less dense than outside of this region.
This means that the macroscopic state of the system is different here 
due to an abrupt change in the heat capacity.
In order to check what is the reason for this transition,
we investigate the pressure distribution in the disk.


\begin{figure}
\epsscale{.80}
\plotone{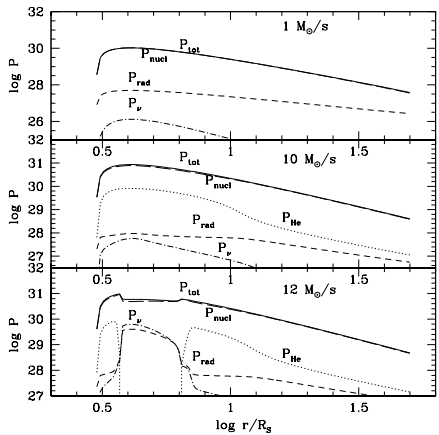}
\caption{The pressure components as a function of the disc radius,
calculated in the stationary solution for
the three accretion rate values:
$\dot M=1~ M_{\odot}$/s (upper panel),
$\dot M=10~ M_{\odot}$/s (middle panel)
 and $\dot M=12~M_{\odot}$/s (bottom panel) .
The total pressure is marked by the solid line, and its components are:
nuclear (gas) pressure (long dashed line), radiation pressure (short dashed line),
helium pressure (dotted line) and neutrino pressure (dot-dashed line).
\label{fig:ptot}}
\end{figure}

The profile of the pressure is shown in Fig.~\ref{fig:ptot}.
The dominant term in the total pressure is due to the nucleons, 
while the radiation
pressure (including electron-positron pairs) 
is always several orders of magnitude smaller.
The neutrino pressure is large in the inner disc, once it gets optically thick to neutrinos
(i.e. for $\dot M \ge 10 M_{\odot}$/s). 
A significant contribution to the pressure
is due to helium at densities high enough for helium to form, 
albeit at temperatures
low enough such that its nuclei are not fully disintegrated.
For the largest accretion rate shown, in the region of 
the temperature excess and inverse density gradient
($3.5-7 R_{\rm S}$), the total pressure
distribution flattens. The helium pressure is now vanishingly small due to
the complete photodisintegration, and the nuclear pressure is slightly
decreased due to the composition change: smaller number density of neutrons and
larger number density of protons.
The substantial contribution to the pressure is now
given by the neutrinos (large optical depths; see below) 
and radiation pressure (increased number
of electron-positron pairs).
From the comparison of Figures \ref{fig:rho}, 
\ref{fig:temp} and the bottom panel of Figure
\ref{fig:ptot}, it can be seen that the total pressure
becomes locally correlated with temperature and anticorrelated with density, thus consituting an
unstable phase.

\begin{figure}
\epsscale{.80}
\plotone{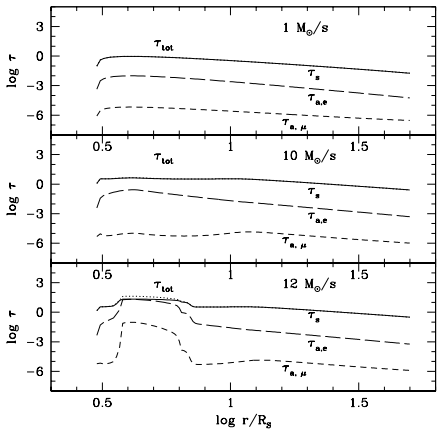}
\caption{The  neutrino optical depths due to 
scattering ($\tau_{\rm s}$, solid line) and absorption ($\tau_{\rm a,e}$ 
for electron neutrinos, long dashed line, and $\tau_{a,\mu}$ for muon 
neutrinos, short dashed line) as a function of radius for 
$\dot M= 1 M_{\odot}/s$ (upper panel),  $\dot M= 10 M_{\odot}/s$ 
(middle panel)  and 
for $\dot M= 12 M_{\odot}/s$ (bottom panel).
The sum of the three quantities is the total optical depth ($\tau_{\rm tot}$, 
dotted line).
\label{fig:tau}}
\end{figure}
In Figure~\ref{fig:tau} we show the neutrino optical depths due to
scattering and absorption. The total optical depth in the outer disc
is typically dominated by scattering processes, while in the inner disc
absorption processes take over  for very high accretion rates. 
For  $\dot M = 1M_{\odot}$/s only
the very inner disk radii have optical depth close to 1. 
 For $\dot M = 12 M_{\odot}$/s, in the radial strip of
 $\sim 3.5-7 R_{\rm S}$ the 
disk is optically thick with
absorptive optical depth  
for electron neutrinos exceeding the scattering term and reaching 
values of the order of  100.

\subsubsection{Composition and Chemical potentials}
\label{sec:chemical}
\begin{figure}
\epsscale{.80}
\plotone{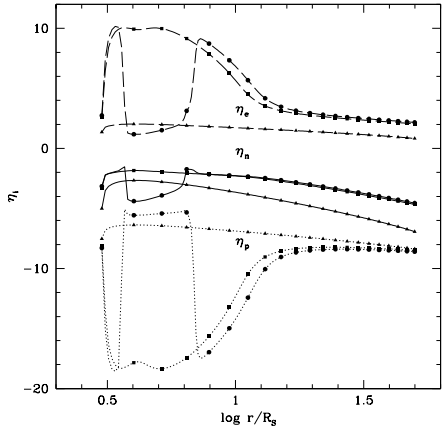}
\caption{The chemical potentials
of neutrons ($\eta_{\rm n}$, solid line), electrons 
($\eta_{\rm e}$, dashed line) 
and protons ($\eta_{\rm p}$, dotted line) as a function of the disc radius,
calculated in the stationary solution.
The accretion rate is
$\dot M=1~ M_{\odot}$/s (triangles),  $\dot M=10~ M_{\odot}$/s (squares)
 and $\dot M=12~ M_{\odot}$/s (circles).
\label{fig:eta}}
\end{figure}

\begin{figure}
\epsscale{.80}
\plotone{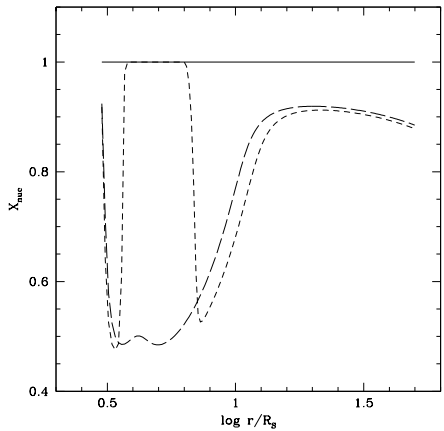}
\caption{The  mass fraction of free nucleons as a 
function of radius for $\dot M= 1 M_{\odot}/s$ (solid line),
 $\dot M= 10 M_{\odot}/s$ (long dashed line)  and $\dot M=12~ M_{\odot}$/s (short dashed line).
\label{fig:xnucrad}}
\end{figure}

\begin{figure}
\epsscale{.80}
\plotone{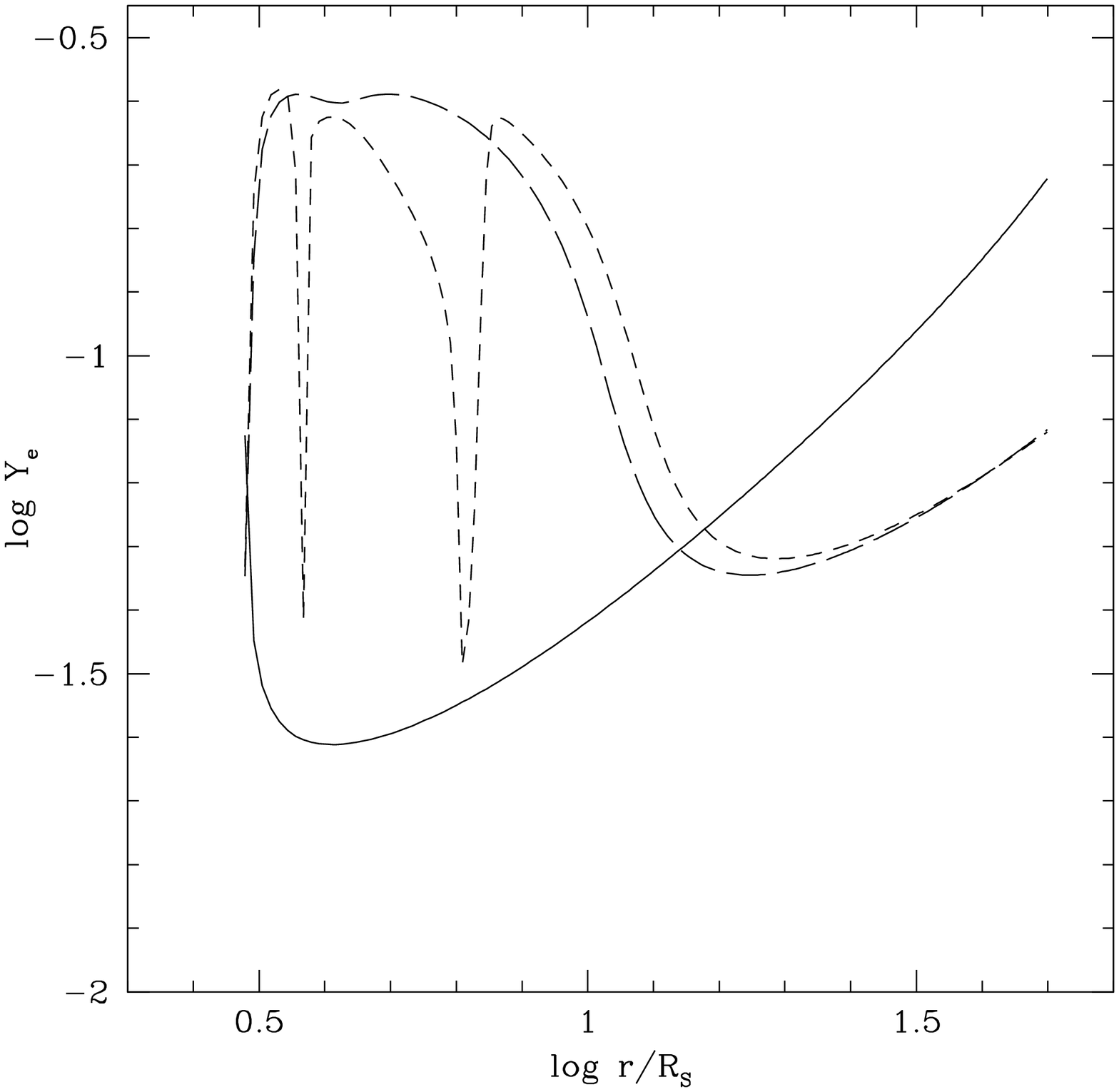}
\caption{The electron fraction as a function of the disc radius,
calculated in the stationary solution.
The accretion rate is
$\dot M=1~ M_{\odot}$/s (solid line), $\dot M=10~ M_{\odot}$/s (long dashed
line)  and $\dot M=12~ M_{\odot}$/s (short dashed line).
\label{fig:ye}}
\end{figure}

In Figure \ref{fig:eta} we show the distribution of the reduced chemical 
potentials of protons, electrons and neutrons throughout the disc.
Reduced electron chemical potentials 
much larger than unity (indicating strong electron degeneracy) 
are found in the inner disc parts
for $\dot M = 10 M_{\odot}$/s and $\dot M = 12 M_{\odot}$/s, 
whereas for 1 $M_{\odot}$/s electrons
are only slightly degenerate.
 For the highest accretion rate,  
the maximum degeneracies correspond to the radius of the local peak in the
density (cf. Fig. \ref{fig:rho}) and the excess of helium number density 
(cf. Fig. 
\ref{fig:xnucrad}). Below this radius, the species become non-degenerate again,
contributing to the increase of the electron fraction (cf. Fig. \ref{fig:ye}).
In Figure \ref{fig:xnucrad} we plot the mass fraction of free nucleons
as a function of radius for $\dot M= 12 M_{\odot}$/s, 10 and $1
M_{\odot}$/s.  As the Figure shows, 
in the outer regions, $X_{\rm nuc}$ increases as the radius decreases,
while the temperature and density increase (Fig. \ref{fig:rho} and Fig. \ref{fig:temp}). 
Consistent with the behavior 
of $Y_{\rm e}$ (Fig.\ref{fig:eos2}), $X_{\rm nuc}$ subsequently turns around 
(decreases) 
at radii where the density is high enough for significant helium
formation. This trend is reversed sharply  for highest accretion rates, 
when the temperature
in the disk is high enough (Fig.\ref{fig:temp}) for helium to be fully dissociated. 
In consequence, the number density of alpha
particles increases at  $\sim 7-12 R_{\rm S}$  and sharply decreases
at lower radii. A similar, but far less pronounced fluctuation in $X_{\rm nuc}$
is seen at smaller radii for the case of  $\dot M= 10 M_{\odot}$/s.
 For smallest accretion rate, $1
M_{\odot}$/s, there is no helium throughout the disk.

In Figure \ref{fig:ye} we show the radial distribution of the electron
fraction throughout the disc for 1, 10  and 12 $M_{\odot}$/s.  For the case of 1
$M_{\odot}$/s (solid line), the electron fraction decreases inward in the disc
as the electrons are captured by protons (in neutronization reactions).  Once
the electrons become non-degenerate, positrons appear, and the positron
capture by neutrons again increases the electron fraction.  For the hotter
plasma (accretion rate of $10~ M_{\odot}$/s, dashed line), consistently
with the behavior discussed for $X_{\rm nuc}$, helium
nuclei form as the density becomes high enough below $\sim 20 R_{\rm S}$
and $Y_{\rm e}$ increases. 
 For the accretion rate of $12~M_{\odot}$/s there is 
the sharp decrease in $Y_{\rm e}$, at 
$\sim 7-8 R_{\rm S}$, due to the sudden dissociation of helium.
As helium is fully photo-dissociated, there is an
almost equal number of neutrons and protons due to the balance of the electron 
 and positron capture. This implies an electron fraction of 0.5.
 At the innermost radius, the temperature and density drop due to
the boundary condition, which affects the behaviour of both $Y_{\rm e}$ and $X_{\rm nuc}$. 

\subsubsection{Cooling and heating rates}
\begin{figure}
\epsscale{.80}
\plotone{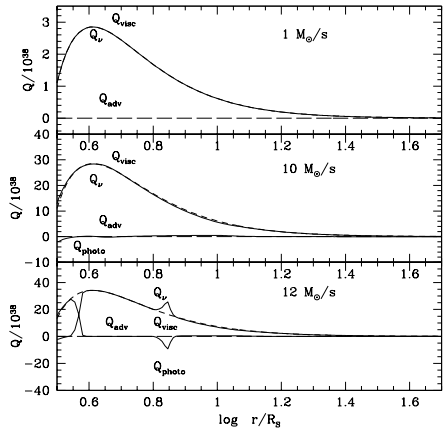}
\caption{The  heating and cooling rates due to 
photodissociation and neutrino emission 
(solid lines) 
 as a function of radius for $\dot M= 1 M_{\odot}/s$ (upper panel),
  $\dot M= 10 M_{\odot}/s$ (middle panel)  and 
for $\dot M= 12 M_{\odot}/s$ (bottom panel).
The other terms are: cooling rate due to advection (long dashed line) 
and viscous heating rate
(short dashed line).
\label{fig:qnu2}}
\end{figure}

In Figure \ref{fig:qnu2} we plot the rates of viscous heating, advection
and  cooling due to neutrino emission and photo-dissociation 
in the stationary disc. 
The accretion rates are $\dot M=1~ M_{\odot}$/s (upper panel),
 $\dot M=10~ M_{\odot}$/s (middle panel),
 and $\dot M=12~ M_{\odot}$/s 
(lower panel). For the highest accretion rates, in the innermost disc the 
neutrino cooling rate decreases substantially with respect to the 
cooling by photodissociacion. 
This is because the neutrinos are trapped in the disc due to a large
opacity 
The smaller the accretion rate, 
the less important is the neutrino trapping effect.
This implies that for an accretion rate of $\le 10~ M_{\odot}$/s neutrinos can 
escape from the innermost disc. 

The 
 advective term is  a couple orders of magnitude smaller than
the other terms.
The photodissociacion term is negligible 
for an accretion rate of $1~
M_{\odot}$/s, since there is no helium in the whole disc, and
$Q_{\rm photo}$ is equal to zero  by definition.
For an accretion rate of $10~ M_{\odot}$/s there is  very little
helium down to about 15-20 $R_{\rm S}$, and
therefore $Q_{\rm photo}$ is  much smaller than other terms. 
 For the accretion rate of $12~ M_{\odot}$/s ,
down to $6-10 
R_{\rm S} $ in the region of the disc of high density and maximum
degeneracy, helium nuclei form.  The nucleosynthesis of alpha
particles leads to the plasma heating instead of cooling, and
therefore the relevant term in the energy balance has a negative
value. Outward, above $\sim 10 R_{\rm S}$, there is some
fraction of helium which can be photo-dissociated, so the cooling term
due to this reaction is also important in the total energy balance.
In the inner region helium is fully dissociated and $Q_{\rm photo}$ is
equal to zero,  increasing again only near the inner boundary
due to the local density increase and decrease of temperature.

\subsection{Stability analysis: instabilities at high-$\dot{M}$}
The disc is thermally unstable 
if $d \log Q^{+}/d \log T > d \log Q^{-} /d \log T$. 
Then any small increase
(decrease) in temperature leads to a heating rate which is more (less)
than the cooling rate, and as a consequence a further increase (decrease)
of the temperature. 
The viscous instability, which appears when
${\partial \dot M \over \partial \Sigma}_{| Q^{+}=Q^{-}} < 0$,
manifests itself in a faster (slower) evolution of an underdense (overdense)
region. The instabilities can be conveniently located in the 
surface density - temperature diagrams, in which the branch of thermal
equilibrium solutions with a negative slope is not only unstable
to the perturbations in the surface density, but it is also thermally
unstable.

\begin{figure}
\epsscale{.80}
\plotone{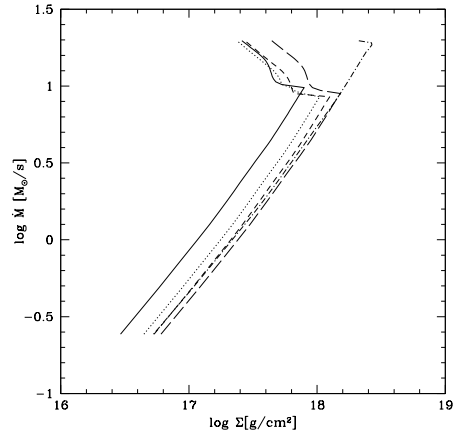}
\caption{The stability curves on the 
accretion rate vs. surface density plane, for several chosen radii in the disc:
$3.39 R_{\rm S}$ (solid line),  $3.81 R_{\rm S}$ (dotted line), $4.25 R_{\rm S}$ 
(short dashed line), $5.19 R_{\rm S}$ (long dashed line)
and $8.60 R_{\rm S}$ (dot-dashed line).
\label{fig:stabcurves}}
\end{figure}

In Figure \ref{fig:stabcurves} we show such stability curves 
for several radii in the disc. The criterion for a viscously stable disc
is generally satisfied throughout the whole disc for   $\dot M \le 10 M_{\odot}/s$.
However, for larger accretion rates, there are unstable branches at
the smallest radii.
For $\dot M = 10 M_{\odot}$/s, the disc becomes unstable  below 5
$R_{\rm S}$, while for $\dot M = 12 M_{\odot}$/s the instability
strip is up to $\sim 7 R_{\rm S}$. 
Here helium is almost completely photodisintegrated while the
electrons and protons become non-degenerate again.
For this high accretion rate, the electron fraction rises inward in
the disc. Under these conditions, the energy balance is affected
leading to the thermal and viscous instability, as demonstrated by the
 stability curves.
This instability will be discussed in more detail in Section \ref{sec:instab}.

\subsection{Time dependent solutions}
\label{sec:lcurves}
\begin{figure}
\epsscale{.80}
\plotone{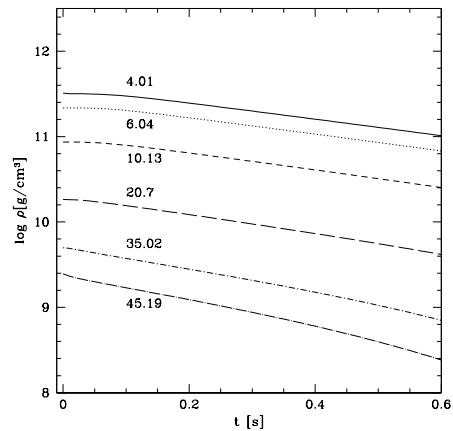}
\caption{The density as a function of time, for several chosen 
disc radii: 4.01, 6.04, 10.13, 20.7, 35.02, and 45.19  $R_{\rm S}$.
The initial accretion rate is
$\dot M=1~ M_{\odot}$/s.
\label{fig:rotime}}
\end{figure}

\begin{figure}
\epsscale{.80}
\plotone{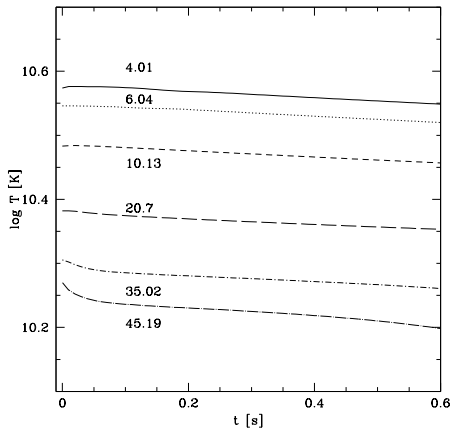}
\caption{The temperature as a function of time, for several chosen 
disc radii: 4.01, 6.04, 10.13, 20.7, 35.02, and 45.19  $R_{\rm S}$.
The initial accretion rate is
$\dot M=1~ M_{\odot}$/s.
\label{fig:ttime}}
\end{figure}

In this Section, we discuss how the temperature, density, electron
fraction and disk luminosity evolve with time.  In Figures
\ref{fig:rotime} and \ref{fig:ttime} we show the time evolution of
density and temperature, when the initial accretion rate is 1
$M_{\odot}$/s.  These quantities exponentially decrease with time:
\begin{equation}
\rho =\rho_{0}(r) \exp (-a t)\;,
\end{equation}
and
\begin{equation}
T =T_{0}(r) \exp (-b t)\;
\end{equation}
where $a \approx 1.9$ and $b \approx 0.085$. The normalization of these 
relations depends on the radius, and for example for $r=6 R_{\rm S}$
it is  $\rho_{0}=2.2\times 10^{11}$ and $T_{0}=3.5\times 10^{10}$.
The exponential behaviour arises from the nature of energy
  equation (45).

\begin{figure}
\epsscale{.80}
\plotone{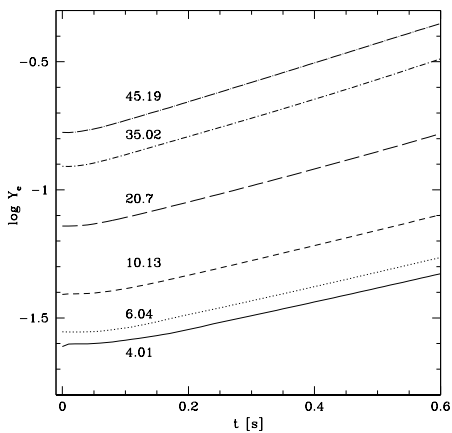}
\caption{The electron fraction as a function of time, for several chosen 
disc radii: 4.01, 6.04, 10.13, 20.7, 35.02, and 45.19 $R_{\rm S}$.
The initial accretion rate is
$\dot M=1~ M_{\odot}$/s.
\label{fig:yetime}}
\end{figure}

In Figure \ref{fig:yetime} we show the electron fraction as a 
function of time for several exemplary radial locations in the disc,
for the disc evolving from a starting accretion rate of $1 M_{\odot}/s$.
The fraction $Y_{\rm e}$ is smaller
in the inner disc radii, while 
outward, the electron fraction is over half an  order of magnitude higher.
Altogether, during the evolution of the system, 
the electron fraction constantly increases with time throughout the disc.

The time-dependent neutrino luminosity of the disc is given by:
\begin{equation}
L_{\nu}(t)=\int_{R_{\rm min}}^{R_{\rm max}} Q^{-}_{\nu}(t) 2\pi r dr
\label{eq:lumthin}
\end{equation} 
where $Q^{-}_{\nu}$ is given by Equation (\ref{eq:qnuthick}).

\begin{figure}
\epsscale{.80}
\plotone{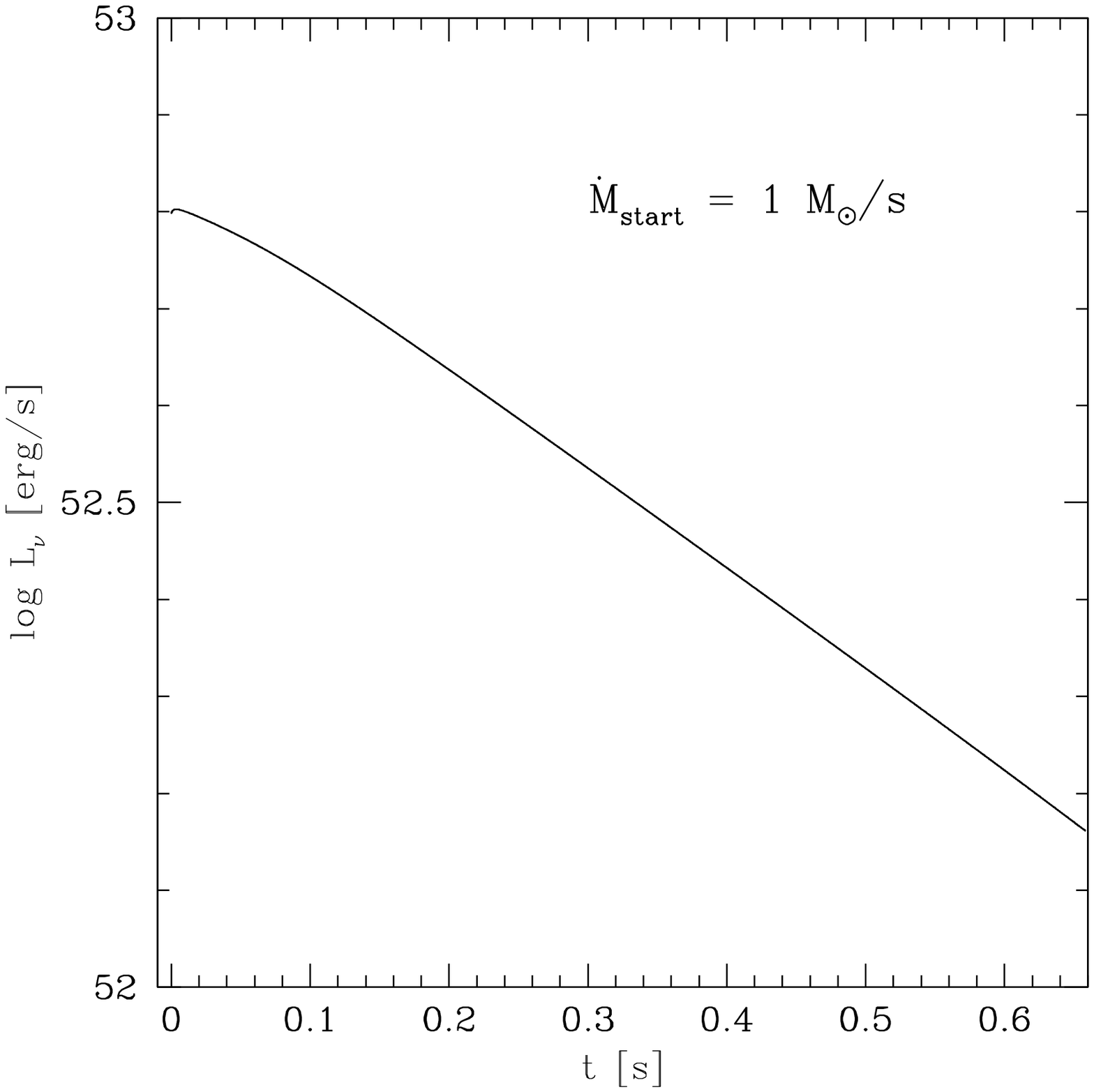}
\caption{The neutrino lightcurve, integrated
over the disc surface.
The initial accretion rate is
$\dot M=1~ M_{\odot}$/s.
\label{fig:alfa}}
\end{figure}

In Figure \ref{fig:alfa} we show an example of such a lightcurve, for
our standard model parameters ($M=3M_{\odot}$, $\alpha = 0.1$), and
$R_{\rm max} = 50 R_{\rm S}$.  The starting accretion rate was $\dot
M_{\rm start}=1 ~M_{\odot}$/s. At this accretion rate neutrinos can
already escape from the accretion disc at the beginning of the
evolution.  For higher initial accretion rates, e.g $10-12 M_{\odot}/s$,
neutrinos are trapped in the innermost disc, and, as a consequence,
the neutrino luminosity is lower at the initial stages of disc
evolution until the accretion rate drops to about $\sim 1
M_{\odot}/s$.  This result is qualitatively similar, albeit it differs
quantitatively, from what was obtained in Di Matteo et al. (2002) and
Janiuk et al. (2004): in those calculations neutrino trapping was far
more substantial even for a 'moderate' accretion rate of $\dot M \ga 1
M_{\odot}$/s. The difference arises from the fact that here we calculate the neutrino
opacities using the $\beta$ reaction efficiencies, self-consistently
with the equation of state.
 
For an accretion rate of $1 M_{\odot}$/s, the solution does not reach
the viscously unstable branch. Initially, the disc contains almost no
alpha particles (cf. Fig. \ref{fig:xnucrad}), which appear later on
during the evolution and cooling of the plasma.  The dynamical balance
between the photodisintegration of helium and nucleosynthesis leads to
an additional non-zero cooling/heating term in the energy equation and
to only small amplitude flickering at the early stages of
time-evolution.

The situation is much more dramatic
when the starting accretion rate is  $12 M_{\odot}$/s. 
In this case a large disc strip is viscously and thermally 
unstable and the most violent instability takes place
around and below $7-12 R_{\rm S}$.

\begin{figure}
\epsscale{.80}
\plotone{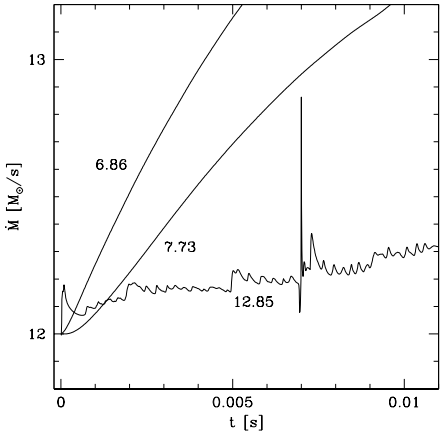}
\caption{The local accretion rate, as a function of time,
for several chosen radial locations in the disc: 6.86, 7.73
and 12.85 $R_{\rm S}$. 
The starting accretion rate is
$\dot M=12~ M_{\odot}$/s.
\label{fig:mdtime}}
\end{figure}
In Figure \ref{fig:mdtime} we show the behavior of the local accretion rate
in the unstable disc, at several chosen locations within the instability
strip.
Near $\sim 12 R_{\rm S}$, the accretion rate varies due to the
large and rapidly changing photodisintegration term (locally, it can become
larger than the neutrino cooling rate). 

This radius corresponds to the largest local value of the density of helium (cf. 
Figure \ref{fig:xnucrad} showing its starting model distribution), which is then being
photodissociated. The photodissociacion process is the cause of the local 
rapid accretion rate changes.
Then, inside from this highly variable
strip, the accretion rate grows too fast to preserve the disc structure.
This kind of behaviour occurs in the locally hotter and less dense region
visible in the starting configuration 
e.g. in Figures \ref{fig:rho} and \ref{fig:temp}, 
between 3.5 and 7 $R_{\rm S}$.
In this region the helium is already totally photodissociated.
Due to the growing accretion rate 
all the material is rapidly 
accreted onto the black hole and the innermost strip of the disc empties.

After the inner strip is destroyed, the outer parts can still accrete onto the
center. As they approach the black hole, their temperature
and density grow and the above  situation can repeat several times, until 
the whole disc is completely broken into rings and destroyed.
These later injections of energy, with timescales dictated by the
viscous timescale of each ring, can produce energy flares following the
main GRB activity. Our results therefore provide another physical mechanism\footnote{In addition
to the gravitational instability in the outer parts of the disk, which was
hinted by the calculations of Di Matteo et al. (2002) and confirmed by those
of Chen \& Beloborodov (2006).}
for the flare model recently proposed by Perna et al. (2006).

\begin{figure}
\epsscale{.80}
\plotone{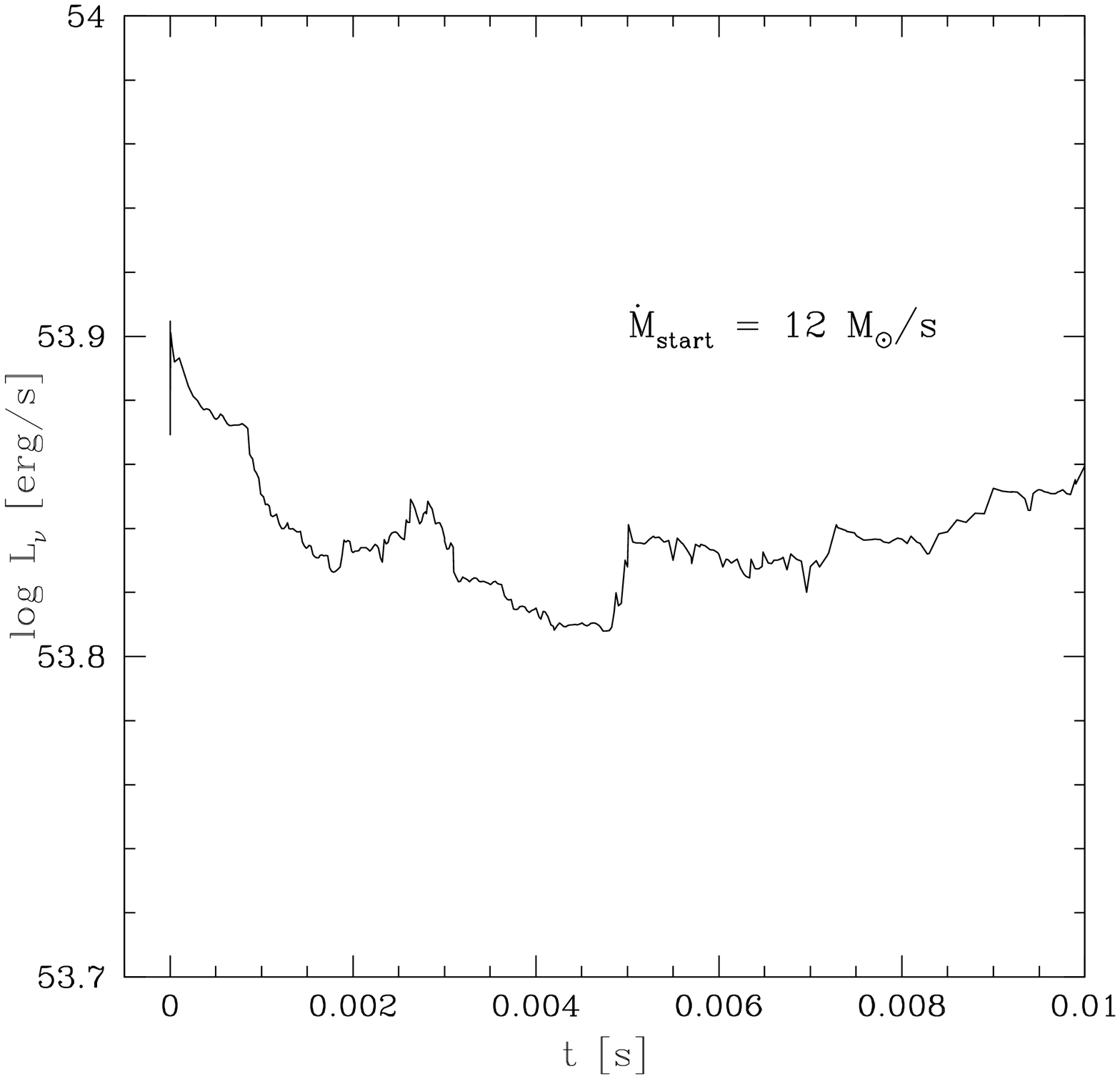}
\caption{The neutrino luminosity, as a function of time 
The starting accretion rate is
$\dot M=12~ M_{\odot}$/s.
\label{fig:lnutime10}}
\end{figure}
In Figure \ref{fig:lnutime10} we show the neutrino lightcurve
of the unstable disc. The instabilities due to 
photodisintegration are reflected in oscillations of variable
amplitude and millisecond timescale. This is of a particular interest
if the neutrino annihilation provides the energy input for GRBs,
however it should be pointed out that the oscillations
appearing in the presented lightcurve have a much smaller amplitude
than the observed gamma ray variability.

\section{Discussion}
\label{sec:diss}

\subsection{The unstable neutrino-opaque disc}
\label{sec:instab}
\begin{figure}
\epsscale{.80}
\plotone{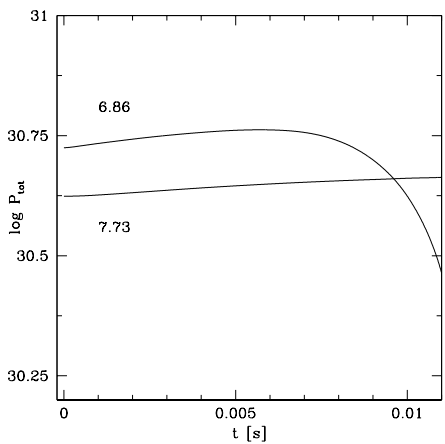}
\caption{The total pressure, as a function of time,
for the chosen radial locations in the disc: 7.73 and
6.86 $R_{\rm S}$. 
The starting accretion rate is
$\dot M=12~ M_{\odot}$/s.
\label{fig:ptime}}
\end{figure}
In our calculations we have shown that, for large accretion 
rates, the accreting torus becomes viscously and thermally unstable.
We now discuss the physical origin of the instability.

The unstable branch appears both in the steady-state solutions and in
the subsequent time-dependent evolutions. In the steady-state case,
for a chosen value of a constant accretion rate, this can be seen for
instance by plotting the radial profiles of density and temperature
(cf. Figs. \ref{fig:rho} and \ref{fig:temp}, where the distinct branch
is found for the innermost radii), as well as by looking at the
stability curves for a range of accretion rates at a chosen disc
radius (cf. Fig. \ref{fig:stabcurves}, where the unstable inner disc
radii exhibit a negative slope in the curve).  In the time-dependent
simulations, the unstable behavior is manifested by the highly
variable accretion rate in certain strips of the disk 
and by the subsequent breaking of the disk inside from 
these variable strips
(cf.
Fig. \ref{fig:mdtime}). The instability arises from the fact that the
accretion rate rises locally too fast to prevent the disc strip from
emptying, as the material is supplied from outer strips at much slower
rate than it is accreted inwards.  The disc evolves unstably on a
viscous timescale, $\tau_{\rm visc} = 1/(\alpha\Omega)\times
(r/H)^{2}$; for the radii shown in Figure \ref{fig:mdtime}, it is
$\tau_{\rm visc} \approx 0.05$ s (note that the disc is rather thick,
$r/H \sim 2.5$, and therefore the viscous and thermal timescales are
close to each other).  Theoretically, in order to find again a stable
solution, the disc would have to increase the local accretion rate up
to about several tens of $M_{\odot}$/s within one viscous
timescale. However, this may not be possible if there is not enough
material in the system to support much higher accretion rates during
such violent oscillations. Therefore the system is unable to be
stabilized and gets broken after a fraction of $\tau_{\rm visc}$ .  In
addition, the dynamical instability is the source of the flickering of
the local accretion rate at the edge of the unstable strip.

Let us now discuss in more detail the physical reason driving this instability.
In the inner part of the disc (below $r\sim 10 R_{\rm S}$ for  $\dot M = 12 
M_{\odot}$/s) there are two important processes, both of which are incorporated
in our equation of state: photodisintegration of helium and neutrino trapping.
As it was already mentioned in Sec. \ref{sec:chemical} and 
can be seen from Fig. \ref{fig:xnucrad}, below  $\sim 7-8 R_{\rm S}$
helium in this disc is completely photodisintegrated.
This part of the disc is also opaque to neutrinos, as we show in 
Fig. \ref{fig:tau}.

These two mechanisms competitively influence the electron fraction in the disc
(cf. Sec. \ref{sec:stabil}, Fig. \ref{fig:tau} and Fig. \ref{fig:ye}). 
Well outside the unstable strip, above $\sim 20 R_{\rm S}$, the electron
fraction smoothly decreases inwards as positrons appear 
because of the neutronization process. Then the scattering optical 
depth for neutrinos becomes $\tau_{\rm s} > 1$, and the electron fraction 
increases again. 
After photodisintegration, the electron fraction decreases significantly from
almost 0.3 to much less than 0.1 due to electron capture. But again,
when the disc becomes optically thick to absorption of electron neutrinos,
the electron fraction gets higher and approaches almost 0.5.

The total pressure of sub-nuclear matter 
(cf. Fig \ref{fig:ptot}) is mainly contributed by electrons,
and therefore it is influenced by the changes in the electron fraction.
In the narrow range of radii  ($6.8<r/R_{\rm S}<7.8$), 
the pressure decreases due to photodisintegration. 
The sudden decrease
of the pressure might drive the dynamical instability. (This picture is
somewhat similar to that of the iron core collapse in the core collapse 
supernova explosions: 
electron capture consumes most of the electrons and makes the EOS
softer, consequently, it triggers the collapse of the iron
core.)
However, the transition from neutrino transparent 
to opaque disc, and the increase of the electron fraction due to the beta
 equilibrium (see also Yuan \& Heyl 2005), 
are the reason for a steeper increase of 
the total pressure of the system.

The same effect can also be observed in the time-dependent plot
(Fig. \ref{fig:ptime}), in which we show the pressure changes in the
characteristic radii of the unstable part of the disc
(cf. Fig. \ref{fig:mdtime}).  The pressure decreases with time up to a
radius  $R= 6.8 R_{\rm S}$, since the temperature and density
gradually drop, as well as the neutrino opacities, so the electron
fraction gets smaller.  Then, in a strip between $\sim$ 6.8 and 12
$R_{\rm S}$, the pressure rises with time: in fact, when $\alpha$
particles appear, the electron fraction rises and matter locally piles
up, thus increasing the pressure.

At the border of these radii the disc breaks up, when the
thermal-viscous instability induces an avalanche-like increase of the
local accretion rate below $\sim 8 R_{\rm S}$.  This happens
because the increase in the pressure causes an excess in the local
energy dissipation rate and the disc heats up, while at adjacent
radii the pressure decreases and heating is insufficient.  The
system tries to compensate these temperature gradients by decreasing/
increasing the temperature in the outer/inner radius, respectively.
But since in the unstable mode of the thermal balance this causes a
further increase/decrease of density, the pressure drops further and
the disc heats up in the outer radius, while cools down in the inner
one.  As long as it cannot find any stable track of evolution, the
emptying of the inner strip continues and finally the whole material
is accreted towards the black hole or blown out.

The radial extent of the unstable part depends on the initial accretion rate
and in our model for  $\dot M = 12 M_{\odot}$/s it is up to $\sim 8 R_{\rm S}$,
 for $\dot M = 10 M_{\odot}$/s it is up to $\sim 5 R_{\rm S}$, while 
for $\dot M = 1 M_{\odot}$/s it is below $\sim 3.5 R_{\rm S}$. 
In the latter 
case,
since the inner radius is located at $\sim 3 R_{\rm S}$, the instability hardly 
affects the disc.
The extension of the instability strip depends also on the mass
of the accreting compact object, and since for lower mass black holes the 
accretion 
disc is generally denser, it reaches a density $\sim 10^{12}$ g/cm$^{3}$
around $\sim 15 R_{\rm S}$.

Above 25-30 $R_{\rm S}$, 
where the plasma is already optically thin and the evolution is stable, 
both the pressure and the accretion rate smoothly drop with time.
The dominant source of cooling of the disc in this region is
the neutrino emission (advective cooling decreases as the disc transits 
from neutrino opaque to transparent).
The photodisintegration term (if non-zero), is usually 
by 1-2 orders of magnitudes smaller than neutrino 
cooling, 
and in the inner disc, up to about 
6.8 $R_{\rm S}$, there are no helium nuclei and the photodisintegration term is 
negligible, while at 7.5 $R_{\rm S}$ 
it has a value of about $Q_{\rm photo} \sim 10^{38}$ erg/s/cm$^{2}$ 
with rapid fluctuations. These fluctuations 
induce the local accretion rate flickering (cf. Fig. \ref{fig:mdtime}),  
on a timescale and amplitude much smaller than for the viscous instability.

\subsection{Comparison with previous work}

The neutrino dominated accretion flow has already been studied in a
number of papers, including both 1-D models and multi-D simulations.
The steady-state 1-D models (e.g Popham et al. 1999; Kohri \&
Mineshige 2002) assumed the disk optically thin to neutrinos, and
neglected photodisintegration cooling.  Di Matteo et al. (2002) took
these two effects into account, and showed that the trapped neutrinos
dominate the pressure in the inner region of the hyperaccreting disc,
however their equation of state did not include the numerical
calculation of chemical equilibrium and did not incorporate the
opacities directly in the EOS iterations (see also the time-dependent
model of Janiuk et al. 2004).  Kohri, Narayan \& Piran (2005)
considered the neutrino opaque disk and the equilibrium between
neutrons and protons and calculated the number densities of species by
numerically integrating their distribution functions.  However, these
authors calculated the gas pressure from the ideal gas approximation,
and neglected the contribution of helium to the pressure.  In all of
these papers the disc occurred to be stable against any kind of
instability.
 
On the other hand, in their recent work, Chen \& Beloborodov (2006) find that
the outskirts of the disk are gravitationally unstable. The approach used by
these authors provides a detailed treatment of the microphysics which is very
similar to ours; however, some differences between our work and theirs must be
crucial to the development of viscous and thermal instabilities. One
difference deals with the approximation made for the treatment of transition
region between the neutrino-opaque and the transparent matter. In our work, we
adopted a gray body model, i.e., we introduced the $b$ factor to describe the
distribution function (c.f. Eq.~10). This assumption is consistent with the
two fluids approximation we have made, which has recently been studied
numerically by Sawyer (2003), and shown to be appropriate for the conditions of
these disks. On the other hand, Chen \& Beloborodov (2006) smoothly connect
the optically thin and thick regimes by means of interpolation. A further
difference lies in the description of the mass fraction of free nucleons. In
this work we use an expression for $X_{\rm nuc}$ developed by Qian et al
(1996), while in Chen \& Beloborodov (2006) $X_{\rm nuc}$ is a function of
$Y_{\rm e}$, which couples the nucleosynthesis to the electron fraction.

In our calculations we reach the range of densities and temperatures
where the nucleons start to become partially degenerate. This is
accompanied by the neutrinos being more and more trapped in the gas
and helium being destroyed by photodissociation. As a result of these
calculations, we found an additional, unstable branch of solutions for
the disc thermal balance.

This supports the recent results of 2-D simulations by Lee,
Ramirez-Ruiz \& Page (2005), who found the disc opaque to neutrinos to
be thermally unstable.  Their simulations showed that large
circulations develop in the accretion flow.  Setiawan, Ruffert and
Janka (2005) found small fluctuations of the accretion rate and
neutrino luminosity on the dynamical timescale, after the 10-20 msec of 
relaxation period (note, that in our calculations we start from the
steady-state disc model at a given accretion rate, thus having no need
for a relaxation to the quasi-steady configuration).  The equation of
state used in their work (see also Janka et al. 1999) is based on the
work of Lattimer \& Swesty (1991).  Given the electron fraction, this
EOS assumes the condition of nuclear statistical equilibrium without
neutrino trapping, but the evolution of the electron fraction is
affected by the asymmetric neutrino emission from the hot and dense
matter, which is called `neutrino leakage scheme'.  The neutrino
leakage scheme focuses on the effects of the neutrino trapping on the
net neutrino emissivities, not on the nuclear statistical equilibrium.
The equation of state used in the work of Rosswog et al. (2004) is
temperature and composition dependent, based on the relativistic mean
field theory (Shen et al. 1998a,b), and the neutrino cooling is
accounted for by the multiflavor scheme (Rosswog \& Liebendoerfer
2003).

In our work, we use an equation of state based on the $\beta$
equilibrium, including the contribution from the trapped neutrino, and
neutrino trapping effects are accounted for by the appropriate
opacities. It should be emphasized that most previous multi-D
simulations neglected the effects of neutrino trapping on the $\beta$
equilibrium, as well as the contribution of the trapped neutrinos to
the thermodynamical properties of the dense matter.  
Another difference between our treatment of the EOS and the
previous numerical simulations is that we include the
cooling of the photodisintegration of helium.  Even though the
original EOS of Lattimer \& Swesty (1991) 
can provide detailed information about the
composition of the dense matter, this information was not considered in order
to keep the table of the EOS as small as possible (see e.g. Ruffert et
al. 1996) just for numerical reasons. In this way, the
disintegration cooling had not been investigated without the
information on the composition.  Our results indicate that
photodisintegration significantly affects the energy balance.

\subsection{Limitations of our model}

We find the thermal-viscous instability
to be an intrinsic property of
the disc for extremely large torus 
densities (about 10$^{12}$g cm$^{-3}$) and high accretion rates 
 ($\dot M \ge 10 M_{\odot}/s$).
This is seen both in the steady-state results
(radial profiles of density and temperature) and in the subsequent time
evolution. 

Thermal and viscous instabilities have been studied in the case of
standard accretion discs around compact objects (Lightman \& Eardley
1974; Pringle 1977; Shakura \& Sunyaev 1976).  Two main physical
processes that lead to disc instabilities were invoked to explain the
time-dependent behavior of various objects: partial ionization of
hydrogen in the discs of Dwarf Novae (e.g. Meyer \& Meyer-Hofmeister
1981; Smak 1984) and domination of radiation pressure in the X-ray
transients (e.g Taam \& Lin 1984).  Such instabilities do not have to
lead to a total disc breakdown, but rather to a limit-cycle behavior,
if only an additional (i.e. upper) stable branch of solutions can be
found. This might be a hot state with a temperature above $10^{4}$ K,
or a slim disc, dominated by advection (Abramowicz et al. 1988).  In
our 1-D calculation the disc in the GRB central engine is not
stabilized but rather breaks down into rings, as no stable solutions
are reached (possibly, for even higher accretion rates again a stable
part near the black hole could be formed - but these extremely high
accretion rates would not be produced by any compact merger scenario).
Therefore, instead of a limit-cycle activity, what we find here are
several dramatic accretion episodes on the viscous timescale.  The
remaining parts of the torus will subsequently accrete and, while
approaching the central black hole, will get hotter and denser,
breaking at $\sim 7 R_{\rm S}$.

Of course, it would be interesting to study whether such a violent
instability would occur also in the 2D or 3D simulations.  This is
indeed likely to be the case, since as the multi-dimensional
simulations of accretion discs show, the instabilities derived first
in 1D are still present in the hydrodynamical simulations of flows
with non-Keplerian velocity fields (e.g. Agol et al. 2001; Turner
2004; Ohsuga 2006). Possibly, the instability region would be located 
at other (larger) radii if the calculations included the vertical
structure of the disc: this is dependent on temperature and density,
which above the meridional plane may be larger than the mean value 
considered in the vertically averaged model.

We need to note that our 1D calculations do not take into account the
possible effects of non-radial velocity components in the fluid.  For
example, the inverse composition gradient that leads to the disk
instability, might be stabilized by rotation (e.g. Begelman \& Meier
1982; Quataert \& Gruzinov 2000).  In the 2-D simulation of Lee et al
(2005) the neutrino opaque disk exhibits circulations in the r-z
direction.  Such meridional circulations are known to be present in
the Keplerian accretion disks (e.g. Siemiginowska 1988), however it is
unclear if they could always provide a stabilizing mechanism for the
thermal-viscous instability.  Possibly, if the nonradial motions of
the flow provided a stronger stabilizing effect, the disk would
exhibit oscillations in the viscous timescale, without breaking,
similarly to the outbursts of Dwarf Nova disks.

The assumption of the $\beta$ equilibrium (justified, as the mixture
of protons, electrons, neutrons and positrons is able to achieve the
equilibrium conditions) might also have an effect on this result, as
the equilibrium conditions reduce the heating and entropy in the gas.
In fact, the $\beta$ equilibrium condition which is satisfied in the
innermost part of a hyperaccreting disc that is optically thick to
neutrinos, is $\mu_{\rm n}=\mu_{\rm p}+\mu_{\rm e}$. Once the disc
becomes transparent in its outer part, this condition is no longer
valid.  Analytically, it has been derived by Yuan (2005) that the
condition for $\beta$ equilibrium in this case is $\mu_{\rm
n}=\mu_{\rm p}+ 2 \mu_{\rm e}$.

\subsection{Observational consequences}

Our findings might be relevant for interpreting some recent 
observations.  The
flickering due to the photodisintegration of 
alpha particles may lead to a variable energy output on small (millisecond) 
timescales.
The consequence of this may be variability in the
gamma ray luminosity, although the changes in the local accretion
rate may be spread by viscous effects 
(in the lightcurve $L_{\nu}(t)$
integrated over the whole surface of the disc, the millisecond variability
is somewhat smeared, and the amplitudes are not very large). 
Therefore the mass accreted by the black hole
may not be varying substantially, while some irregularity in the overall 
outflow could help produce internal shocks.
  
The thermal-viscous instability, if accompanied by the disc breaking,
may lead to the several episodic accretion events and several
re-brightenings of the central engine on longer timescales, possibly detected
in the later stages of the evolution.
A similar kind of a long-term activity is possible also 
if the disk was not completely broken, but
exhibited some large accretion rate fluctuations on the viscous timescale.

\acknowledgements
We thank Bo\.zena Czerny, Pawe{\l} Haensel  and Daniel Proga
for helpful discussions.
We also thank the anonymous referee for detailed reports which helped
us to improve our model and its presentation.
This work was supported in part by 
grant 1P03D 00829 of the Polish State Committee for Scientific Research and
by NASA under grant NNG06GA80G.
Y.-F. Y. is partially supported by Program for New Century Excellent Talents 
in University, and the National Natural Science Foundation (10233030,10573016).
RP acknowledges support from NASA under grant NNG05GH55G, and
from the NSF under grant AST~0507571.

\appendix

\section{Appendix}
The neutrino absorption and production rates in the beta processes 
for all participating particles at arbitrary degeneracy have been
obtained in the previous works (Reddy, Prakash \& Lattimer 1998;
Yuan 2005). In the subnuclear dense matter with high temperatures,
the nucleons are generally nondegenerate, therefore,  
the transition reaction rates from neutrons to protons and 
from protons to neutrons can be simplified as follows:
\begin{eqnarray}
\Gamma_{p + e^{-} \rightarrow n + \nu_e} 
&=&\frac{1}{2\pi^3}|M|^2 \int_Q^{\infty}dE_e
        E_ep_e(E_e-Q)^2f_e(1-b_ef_{\nu_e}), \\
\Gamma_{p + e^{-} \leftarrow n + \nu_e} 
&=&\frac{1}{2\pi^3}|M|^2 \int_Q^{\infty}dE_e
        E_ep_e(E_e-Q)^2(1-f_e)b_ef_{\nu_e}, \\
\Gamma_{n+e^{+}\rightarrow p+\bar\nu_e} 
&=&\frac{1}{2\pi^3}|M|^2 \int_{m_e}^{\infty}dE_e
        E_ep_e(E_e+Q)^2f_{e^+}(1-b_ef_{\bar\nu_e}), \\
\Gamma_{n+e^{+}\leftarrow p+\bar\nu_e} 
&=&\frac{1}{2\pi^3}|M|^2 \int_{m_e}^{\infty}dE_e
        E_ep_e(E_e+Q)^2(1-f_{e^+})b_ef_{\bar\nu_e}, \\
\Gamma_{n\rightarrow p+e^{-}+\bar\nu_e} 
&=&\frac{1}{2\pi^3}|M|^2 \int_{m_e}^QdE_e
        E_ep_e(Q-E_e)^2 (1-f_{e})(1-b_ef_{\bar\nu_e}), \\
\Gamma_{n\leftarrow p+e^{-}+\bar\nu_e} 
&=&\frac{1}{2\pi^3}|M|^2 \int_{m_e}^QdE_e
        E_ep_e(Q-E_e)^2 f_{e}b_ef_{\bar\nu_e}. 
\end{eqnarray}
Here $Q=(m_{n}-m_{p})c^{2}$, $|M|^2$ is the averaged transition 
rate which depends on the initial and final states of all participating
particles, for nonrelativistic noninteracting nucleons,
$|M|^{2}=G_{\rm F}^{2}\cos^{2}\theta_{\rm C}(1+3g_{\rm A}^{2})$,
here $G_{\rm F} \simeq 1.436 \times 10^{-49}~{\rm erg}~ 
{\rm cm}^{3}$ is the Fermi weak interaction constant,
$\theta_{\rm C}$ $(\sin\theta_{\rm C}=0.231)$
is the Cabibbo angle, and
$g_{\rm A}=1.26$ is the axial-vector coupling constant.
$f_{e,\nu_e}$ are the distribution functions for electrons and
neutrinos, respectively. The ``chemical potential" of neutrinos is 
generally assumed to be zero. The factor $b_e$ reflects the
percentage of the partially trapped neutrinos. When neutrinos
completely trapped, $b_e=1$. 

The corresponding neutrino emissivities for the URCA reactions are
given by:
\begin{eqnarray}
q_{p+e^{-}\to n+\nu_{e}}  
&=&\frac{1}{2\pi^3}|M|^2 \int_Q^{\infty}dE_e
        E_ep_e(E_e-Q)^3f_e(1-b_ef_{\nu_e}), \\
q_{n+e^{+}\to p+\bar\nu_{e}} 
&=&\frac{1}{2\pi^3}|M|^2 \int_{m_e}^{\infty}dE_e
        E_ep_e(E_e+Q)^3f_{e^+}(1-b_ef_{\bar\nu_e}), \\
q_{n\to p+e^{-}+\bar\nu_{e}} 
&=&\frac{1}{2\pi^3}|M|^2 \int_{m_e}^QdE_e
        E_ep_e(Q-E_e)^3 (1-f_{e})(1-b_ef_{\bar\nu_e}).
\end{eqnarray}

The emissivities due to the electron-positron pair annihilation, 
following the notation of Yakovlev et al (2001), is written as:
\begin{eqnarray}
    q_{e^-+e^+\to \nu_i+\bar\nu_i} & = & {Q_c \over 36 \pi} \,
    \left\{ C_{+\nu_i}^2 \left[ 8(\Phi_1 U_2 + \Phi_2 U_1)
           - 2(\Phi_{-1} U_2 + \Phi_2 U_{-1})
           + 7(\Phi_0 U_1 + \Phi_1 U_0) \right. \right.
\nonumber \\        
          & & + \; \left. \left. 5 (\Phi_0 U_{-1}+ \Phi_{-1} U_0) \right]
           +   9 C_{-\nu_i}^2 \left[\Phi_0(U_1+U_{-1}) +
           (\Phi_{-1} + \Phi_1)U_0 \right] \right\},
\end{eqnarray}
where
\begin{equation}
   Q_c= { G_{\rm F}^2 \over \hbar} \,
        \left( m_e c \over \hbar \right)^9
      = 1.023 \times 10^{23} \;\;{\rm erg \; cm^{-3} \; s^{-1}},
\end{equation}
$C_{+\nu_i}=C_{V_{i}}^2+C_{A_{i}}^2$ and 
$C_{-\nu_i}=C_{V_{i}}^2-C_{A_{i}}^2$, here $C_{V_{i}}$
and $C_{A_{i}}$ are the vector and axial-vector constants for
neutrinos ($C_{Ve}=2 \sin^2 \theta_{\rm C} + 0.5$,
$C_{Ae}=0.5$, $C_{V\mu}=C_{V\tau}= 2 \sin^2 \theta_{\rm C} - 0.5$ and
$C_{A\mu}=C_{A\tau}= -0.5$). 
The dimensionless functions $U_k$ and $\Phi_k$ ($k$= $-1$, 0, 1, 2)
in the above equation can be expressed in terms of the Fermi-Dirac
functions:
\begin{eqnarray}
U_{-1}&=&\frac{\sqrt{2}}{\pi^2}\beta^{3/2}F_{1/2}(\eta_e,\beta_e) \\
U_{0}&=&\frac{\sqrt{2}}{\pi^2}\beta^{3/2}[F_{1/2}(\eta_e,\beta_e)
	+\beta_e F_{3/2}(\eta_e,\beta_e)] \\
U_{1}&=&\frac{\sqrt{2}}{\pi^2}\beta^{3/2}[F_{1/2}(\eta_e,\beta_e)
	+2\beta_e F_{3/2}(\eta_e,\beta_e)+\beta_e^2 F_{5/2}(\eta_e,\beta_e)]\\
U_{2}&=&\frac{\sqrt{2}}{\pi^2}\beta^{3/2}[F_{1/2}(\eta_e,\beta_e)
	+3\beta_e F_{3/2}(\eta_e,\beta_e)+3\beta_e^2 F_{5/2}(\eta_e,\beta_e)
	+\beta_e^3 F_{7/2}(\eta_e,\beta_e)].
\end{eqnarray}
Replacing $\eta_e$ with $\eta_{e^+}$ in $U_k$, we get the corresponding
expressions for $\Phi_k$.

\end{document}